\documentclass[acmlarge]{acmart}

\AtBeginDocument{%
  \providecommand\BibTeX{{%
    \normalfont B\kern-0.5em{\scshape i\kern-0.25em b}\kern-0.8em\TeX}}}

\setcopyright{acmcopyright}
\copyrightyear{2023}
\acmYear{2023}
\acmDOI{X}

\let\oldFootnote\footnote
\newcommand\nextToken\relax
\renewcommand\footnote[1]{%
    \oldFootnote{#1}\futurelet\nextToken\isFootnote}
\newcommand\isFootnote{%
    \ifx\footnote\nextToken\textsuperscript{,}\fi}

\usepackage{graphicx}
\usepackage{multirow}

\usepackage{subcaption}

\definecolor{mypink}{HTML}{e72a8a}

\usepackage{tcolorbox}
\newtcolorbox{mybox}{colback=mypink,colframe=mypink}

\usepackage{mdframed}
\newmdenv[
  topline=false,
  bottomline=false,
  skipabove=\topsep,
  skipbelow=\topsep,
  linecolor=mypink,
]{siderules}

\usepackage{enumitem}

\usepackage{xcolor}

\newcommand{\mrev}[1]{\textcolor[rgb]{0.00,0.00,0.00}{#1}}
\newcommand{\rev}[1]{\textcolor[rgb]{0.00,0.00,0.00}{#1}}
\definecolor{outcolor}{rgb}{0.8,0.8,0.8}

\definecolor{myboxcolor}{rgb}{0.9,0.9,0.9}
\newtcolorbox{mybox1}{colback=myboxcolor,colframe=myboxcolor}

\begin{document}
\title{Uncovering Bias in Personal Informatics}

\author{Sofia Yfantidou}
\email{syfantid@csd.auth.gr}
\orcid{0000-0002-5629-3493}
\author{Pavlos Sermpezis}
\email{sermpezis@csd.auth.gr}
\author{Athena Vakali}
\email{avakali@csd.auth.gr}
\affiliation{%
  \institution{Aristotle University of Thessaloniki}
  \city{Thessaloniki}
  \country{Greece}
  \postcode{54124}
  }
\author{Ricardo Baeza-Yates}
\affiliation{%
  \institution{Institute for Experiential AI, Northeastern University}
  \city{San Jose}
  \country{United States}
  \postcode{CA 95113}
  }

\renewcommand{\shortauthors}{Yfantidou et al.}

\begin{abstract}
Personal informatics (PI) systems, powered by smartphones and wearables, enable people to lead healthier lifestyles by providing meaningful and actionable insights that break down barriers between users and their health information. Today, such systems are used by billions of users for monitoring not only physical activity and sleep but also vital signs and women's and heart health, among others. %
Despite their widespread usage, the processing of sensitive PI data may suffer from biases, which may entail practical and ethical implications. In this work, we present the first comprehensive empirical and analytical study of bias in PI systems, including biases in raw data and in the entire machine learning life cycle. We use the most detailed framework to date for exploring the different sources of bias and find that biases exist both in the data generation and the model learning and implementation streams. According to our results, the most affected minority groups are users with health issues, such as diabetes, joint issues, and hypertension, and female users, whose data biases are propagated or even amplified by learning models, while intersectional biases can also be observed.
\end{abstract}

\ccsdesc[500]{Human-centered computing~Ubiquitous and mobile computing}
\ccsdesc[300]{Applied computing~Consumer health}
\ccsdesc[300]{Computing methodologies~Artificial intelligence}
\ccsdesc[300]{Social and professional topics~Codes of ethics}

\keywords{Personal informatics, machine learning, bias, fairness, ubiquitous computing, sensing data, digital biomarkers.}

\maketitle

\section{Introduction}\label{introduction}
Ubiquitous technologies, such as smartphones \rev{and} wearables, are an integral part of our lives today \cite{mobilestatisticsreport,statistaGlobalConnected}. 
Their proliferation has given rise to Personal Informatics (PI), namely a class of systems that ``help people collect personally relevant information for the purpose of self-reflection and gaining self-knowledge'' \cite{li2010stage}. Such systems enable people to keep track of their productivity \cite{kim2016timeaware}, finances \cite{kaye2014money}, and learning \cite{garbett2018thinkactive}. Yet, tracking various aspects of physical and mental health is particularly prevalent \cite{epstein2020mapping}.
PI \rev{systems} can continuously and unobtrusively measure and collect physiological and behavioral data, namely, ``digital biomarkers'', from users through integrated sensors. Digital biomarkers contain an uncanny amount of personal information. Even the coarser behavioral biomarkers acquired from consumer wearables (\rev{e.g.,} steps, calories) strongly correlate to a person's gender, height, and weight \cite{thomas2022fitbit}, while signals of finer granularity (\rev{e.g.,} accelerometer and heart rate), can predict variables associated with an individual's physical health, fitness, and demographics \cite{10.1145/3450439.3451863}. 

\rev{At the same time, consumer smartphones and wearables are now packed with an increasing number of advanced health tracking features, innovating in personal health, research, and care \cite{apple2022health}. 
Flagship consumer wearable algorithms ---some approved by the US Food and Drug Administration--- can now identify signs of atrial fibrillation (AFib) through electrocardiogram (ECG) or photoplethysmography (PPG) signals \cite{fitbitIrregularRhythm}. On a different note, newly released watches introduce novel cycle-tracking functionality, including fertility prediction and notifications, using logged period data and temperature measurements \cite{appleAppleEmpowering}. Mobility features include fall and crash detection through accelerometer and gyroscope measurements and notification of emergency services. Similarly, newly integrated blood oxygen (SpO2) sensors on wearable rings can provide indicators and warn users about potential sleep apnea or lung diseases \cite{ouraringOuraBlood}. It is evident that consumer smartphones and wearables have moved beyond step counts, marking a rapid transition towards mHealth .}

However, the prevalent PI adoption embeds important challenges due to the questionable transparency and unexplored biases in the systems' algorithms. Contrary to the common belief that algorithmic decisions are objective by definition, a machine learning (ML) model may be inherently unfair by learning, preserving, or even amplifying historical biases existent in the data \cite{pessach2022review}. 
\rev{Unfortunately, real-world cases of unfair ML models are abundant even within the ubiquitous computing community. For example, neural network algorithms trained to classify skin lesions were found to exhibit lower diagnostic accuracy in black patients \cite{kamulegeya2019using}. Moreover, racial bias has been identified in health sensors such as oximeters, which were primarily tested on white populations, resulting in the misclassification of people of color~\cite{sjoding2020racial}. Given the growing potential of PI devices in mHealth, imagine the ethical and social implications if an AFib detection algorithm exhibited bias against a specific race or if a fertility prediction algorithm was biased against women in developing countries.}

Despite this growing interest in ML fairness, a focused emphasis on the requirements of unbiased PI systems in mHealth settings is lacking \rev{\cite{ahmad2020fairness,yfantidou2023beyond}}. %
Yet, PI systems are deployed in high-stakes health applications, while their input data modality\rev{, i.e., personal sensitive data,} makes them susceptible to propagating bias. Thus, exploring biases within these systems is critical to raise awareness regarding mitigating and regulatory actions required to avert potential negative consequences. 
This need is further highlighted by \rev{the} differences \rev{between PI and} other domains, such as facial or speech recognition:
\begin{itemize}[leftmargin=*,nosep]
    \item \textbf{\textit{The digital divide as a barrier of entry:}} To contribute data to an image or voice dataset, users do not need any \rev{technical} knowledge or niche device. However, to contribute to a PI dataset, users face significant ``entry barriers'' in terms of digital capacity or device ownership, creating new-found \textit{representation biases}.
    
    \item \textbf{\textit{Emerging technologies accuracy:}} Facial or speech recognition devices, e.g., camera or voice recorder, are mature technologies \rev{with high accuracy}. On the contrary, emerging PI devices' accuracy significantly varies across models, creating unexplored \textit{measurement biases} and discrepancies \rev{across} user segments.
    
    \item \textbf{\textit{Complex nature of data:}} It may be \rev{straightforward} to identify biases in terms of skin \rev{tone}, \rev{nationality} and gender \rev{in facial or speech recognition}. Yet, identifying biases in digital biomarkers, \rev{e.g., sensor data,} \rev{is complicated}. Biases in PI data can remain hidden and be further propagated or even amplified in ML models.
\end{itemize}

Motivated by these \rev{differences} and \rev{research} gap, we present the first comprehensive study on bias in PI: We adopt the most complete framework to date for understanding sources of harm in the ML life cycle~\cite{suresh2021framework} (\S\ref{configuration}), and explore biases in the data generation (\S\ref{dataBias}) and model \rev{building} and implementation (\S\ref{modelBiases}) streams. 
\rev{Ultimately,} we \rev{apply our methodology} in \rev{the largest} real-world PI dataset \rev{to date}, \rev{while providing preliminary indications of generalizability through differing datasets \mrev{and use cases} (\S\ref{generalizability}), and we offer recommendations for bias mitigation (\S\ref{discussion})}.
Specifically, our research questions (RQs) and the respective contributions are as follows:
\begin{enumerate}[leftmargin=*,nosep]
  
    \item \textbf{\textit{Are PI data susceptible to biases?}} \rev{We examine whether ubiquitous digital biomarkers are subject to \textit{historical}, \textit{representation}, and \textit{measurement} biases. To demonstrate our point, we analyze the MyHeart Counts dataset \cite{hershman2019physical}, comprising} physical activity, fitness, sleep, and cardiovascular health data for 50K participants across the United States \rev{(US)}. Our results \rev{reveal} biases across all \rev{stages} of the data generation stream\rev{, highlighting the need for careful usage of} PI datasets, in general, and the MyHeart Counts data, in particular.     
    \item \textbf{\textit{Do ML models inherit PI data biases?}} We examine whether biases \rev{inherent} in PI data \rev{persist during modeling}. %
    Specifically, we assess \rev{various learning and personalization models} for \textit{aggregation}, \textit{learning}, and \textit{deployment} biases. \rev{Consistent with prior research}~\cite{paviglianiti2020vital}, our findings indicate that data biases are propagated to learning models, \rev{particularly} for user groups \rev{with intersecting identities}. They are also significantly amplified in their personalized counterparts, prompting further exploration of personalization trade-offs.   
    \item \textbf{\textit{\rev{Do} synthetic benchmarks hide the imperfect nature of PI?}} We explore whether ``perfect'' synthetic benchmark datasets can hide PI data and model ``imperfections'' and biases during evaluation. Specifically, we compare a random benchmark, representative of our data, with one designed to achieve demographic parity for evaluation biases. Our findings highlight the importance of establishing PI benchmarks that are representative of the intended target populations to avoid deploying models with unidentified biases.
\end{enumerate}

\section{\rev{Background and Related Work}}\label{related-work}
\rev{This section provides the necessary background and delimits the scope of this work, while \rev{discussing} relevant literature on the conceptual space of fairness in PI for mHealth.}
\subsection{\rev{Bias and Fairness in PI Definitions}}
\rev{Bias in ML is a potential source of unfairness that can lead to harmful consequences, such as discrimination. In terms of algorithms, bias can be defined as ``a systematic error or an unexpected tendency to favor one outcome over another''  \cite{mehrabi2021survey}. The term can also refer to an algorithm's undesired dependence on specific data attributes that may be linked to a demographic group, e.g., based on gender, race, or religion \cite{fletcher2021addressing}. While bias is related to fairness, it is important to note that algorithmic bias is distinct from ethics. It is simply a mathematical and statistical consequence of an algorithm, including the data used, the logic itself and the user interaction feedback-loop, making it fully quantifiable \cite{baeza2018bias}}.

\rev{Unlike bias, the fairness of an ML model is judged against a set of legal or ethical principles, which are subject to the local government and culture.
The Fairness, Accountability, and Transparency in Machine Learning (FAccT/ML) community defines fairness as a principle that ``ensures that algorithmic decisions do not create discriminatory or unjust impacts when comparing across different demographics (e.g., race, sex, etc.)'' \cite{awwad2020exploring}. Fairness is an inexorably subjective and context-dependent notion and incorporates different metrics for different definitions, some of which are even mutually incompatible \cite{friedler2021possibility}.}
\rev{For example, drawing from our use case (\S\ref{useCase}) and as per relevant literature \cite{world2019global}, women tend to perform overall less physical activity compared to their male counterparts. Hence, a PI goal-setting algorithm could either give females lower goals because they historically perform less physical activity or give females equal high goals despite historic differences to encourage behavior change. Fairness in this context is subjective and dependent on the viewpoint. However, algorithmic bias is objective and can be identified regardless of fairness considerations. For instance, the first algorithm described would be biased against women since it assigns them fewer high-activity goals compared to men.} 

\rev{Bias can (but does not always) result in discrimination. We consider systems fairer if they are less biased, but building ML systems without bias is practically difficult and possibly infeasible. However, quantifying and mitigating bias is an attainable and important step toward building fairer ML systems. Hence, our work aims to unveil and quantify biases in the PI life cycle without the subjective element of personal fairness perspectives. Note that we might still use the term ``fairness'' in the paper, when we refer to certain standard terminology, e.g., ``fairness through unawareness'' or ``fairness metrics''.}

\subsection{\rev{Bias and Fairness in PI for Health and Well-being Literature}}
With the widespread adoption of intelligent systems and applications in our everyday lives, accounting \rev{for data and model biases} has gained significant traction in designing and deploying systems. Specifically, \rev{these notions} have been studied extensively in domains such as natural language processing \cite{bolukbasi2016man,frey2020artificial}, recommender systems \cite{leonhardt2018user,wang2022survey,wang2020faircharge}, and computer vision \cite{buolamwini2018gender,wang2020towards}. Yet, evidence for \rev{biases} in the PI setting is lacking.
Closer to PI, fairness research in healthcare is still in its infancy \cite{feng2022fair}. The digitization of medical data has enabled the scientific community to collect large amounts of heterogeneous, multi-modal data and develop ML algorithms for various medical tasks. During the process, various limitations have been uncovered based on the three most prominent data types: medical image data, structured electronic health record (EHR) data, and textual data.

Medical imaging has been the most widely used data source for ML in healthcare, and biases in them have received attention \cite{jiang2017artificial}.
For example, \citet{larrazabal2020gender} utilizes two commonly used X-ray image datasets to diagnose various chest diseases under different gender imbalance conditions and showcase that the minority gender group systematically performs worse than the majority gender group. Similarly, according to \citet{adamson2018machine}, relying on ML for skin cancer screening may exacerbate potential racial disparities in dermatology.

On a different note, EHR systems store multi-modal, heterogeneous patient data, such as demographics, diagnoses, and clinical records, and have been used for various tasks, such as medical concept extraction, mortality prediction, and disease inference. Regarding EHR data fairness, \citet{meng2021mimic} identify race-level differences in the predictions of neural network models on the MIMIC-IV dataset \cite{johnson2020mimic}, with Black and Hispanic patients being less likely to receive interventions or receiving interventions of shorter average duration. Similarly, \citet{roosli2022peeking} reveals a strong class imbalance problem and significant fairness concerns for Black and publicly insured ICU patients in the same dataset. 

Concerning textual EHR data, \citet{chen2019can} examines clinical and psychiatric notes to predict intensive care unit mortality and 30-day psychiatric readmission. Their analysis reveals differences in prediction accuracy, and biases are present in terms of gender and insurance type for mortality prediction and insurance policy for psychiatric 30-day readmission. Within the same scope, \citet{zhang2020hurtful} train deep embedding models on medical notes from the MIMIC-III database \cite{johnson2016mimic}, and find that classifiers trained from their embeddings exhibit statistically significant differences in performance, often favoring the majority group regarding gender, language, ethnicity, and insurance status.

Yet, despite the emerging research on \rev{biases} in healthcare, its proximity to PI, and the widespread adoption of PI technologies, biases in PI have been barely explored. 
\citet{paviglianiti2020vital} reported \rev{gender} biases in digital biomarkers using Vital-ECG, a wearable smart device that collects electrocardiogram and plethysmogram signals. Still, their study is limited to quantifying learning bias, far from a complete study of bias in the PI life cycle. \rev{On the contrary, inspired by relevant works in related domains \cite{hutiri2022bias,hutiri2022tiny}, our work aims to raise awareness and set up a systematic approach to comprehensively analyze data and model biases in PI systems, highlighting the multiple facets of bias that may affect system fairness.}

\section{Personal Informatics Biases: Setting \& Configuration}\label{configuration}
In this section, we discuss frameworks \rev{capturing biases} (Section~\ref{framework}) and our use case configuration, \rev{acting} as a starting point for our investigation (Section~\ref{useCase}).

\begin{figure}[t]
\centering
\begin{subfigure}[b]{0.9\textwidth}
   \includegraphics[width=.9\linewidth]{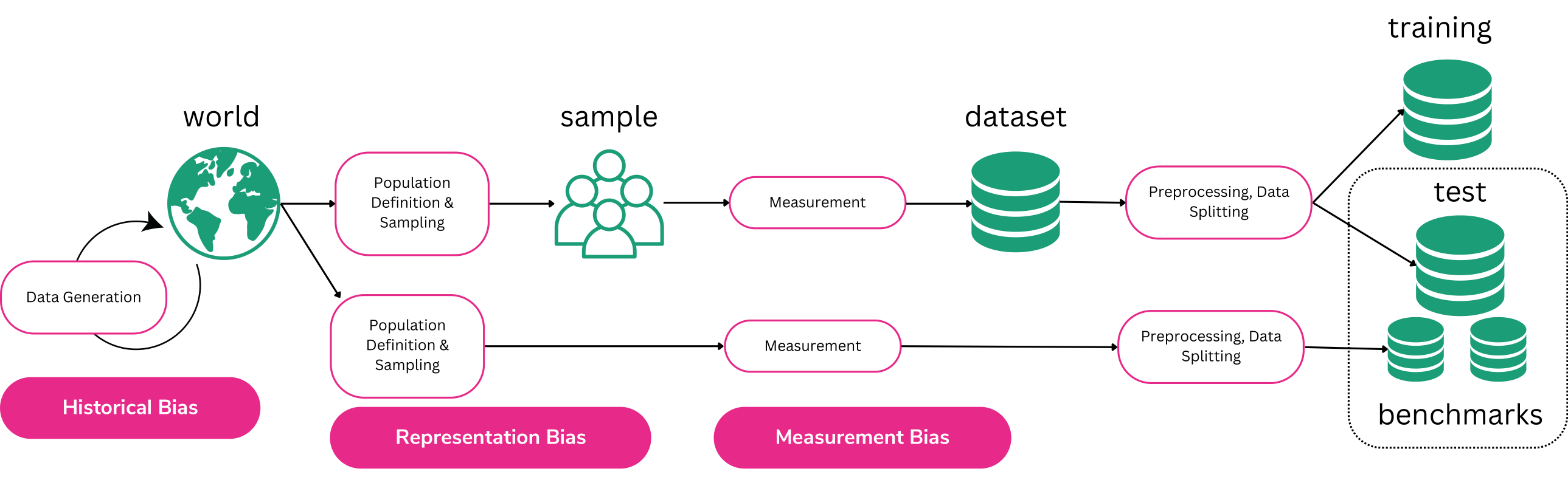}
   \caption{Sources of harm in the data generation stream.\label{fig:biasData}}
\end{subfigure}
\begin{subfigure}[b]{0.9\textwidth}
   \includegraphics[width=.9\linewidth]{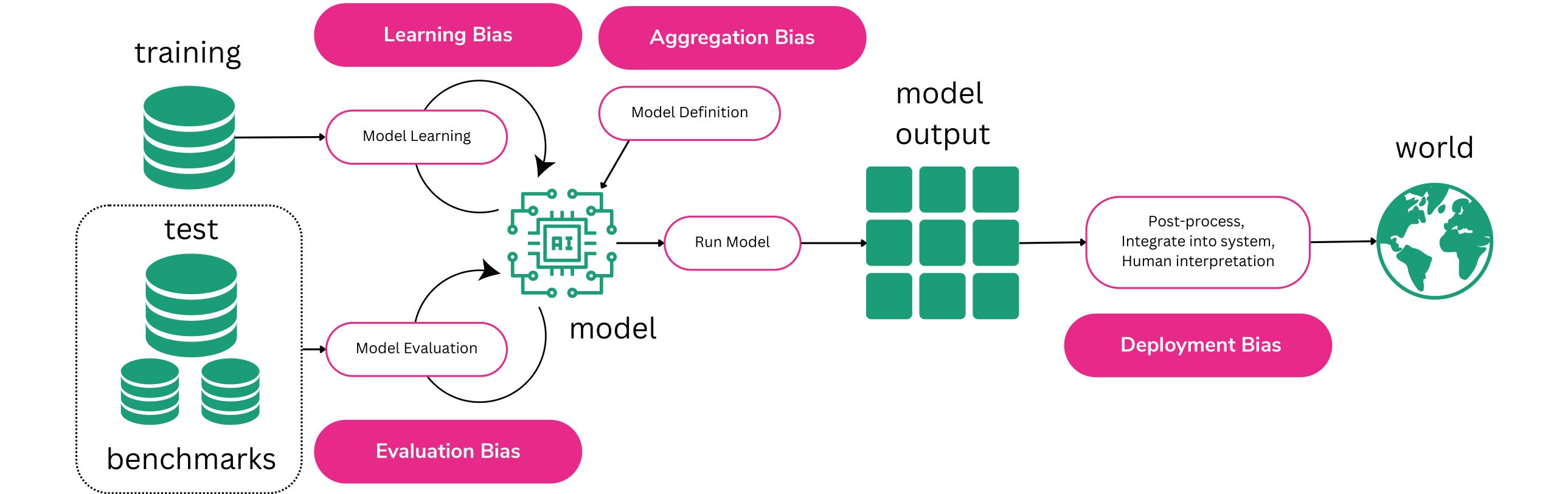}
   \caption{Sources of harm in the model building and implementation stream.\label{fig:biasModel}}
\end{subfigure}
\caption{Sources of harm in the data (top) and model building and implementation (bottom) streams \cite{suresh2021framework}. The training, test, and benchmark sets are common across figures.}
\end{figure}

\subsection{Sources of Bias in the Machine Learning Life Cycle Framework\label{framework}}
\rev{There exist various frameworks for capturing bias in ML applications, e.g., in the context of the Web \cite{baeza2018bias}, autonomous systems \cite{danks2017algorithmic} or crowdsourced labeling \cite{barbosa2019rehumanized}. However, most tend to be domain-specific. Yet, \citet{suresh2021framework} has introduced a framework for understanding sources of harm throughout the ML life cycle, independent of the application domain. Given its generic and comprehensive nature and, thus, suitability for the PI use case, we consider this as a basis for our study.} \rev{As per \cite{suresh2021framework}}, the ML life cycle consists of two streams, the \textit{data generation stream} and the \textit{model building and implementation stream}, containing seven sources of bias-related harms, \rev{as} shown in Figures \ref{fig:biasData} and \ref{fig:biasModel}, respectively, \rev{and defined} below:
\begin{itemize}[leftmargin=*,nosep]
    \item \textit{Historical biases} can occur even if the data are flawlessly sampled by reflecting real-world biases against one or more groups of people. For example, gender gaps in certain \rev{fields can result in} language models \rev{linking certain job-related terms}, such as nurse or programmer, with \rev{female or male descriptors}, respectively \cite{bolukbasi2016man}.
    \item \textit{Representation biases} can occur when sampling methods lead to underrepresenting population segments. For example, in popular image datasets, \rev{skewed towards} the \rev{US} or Europe, \rev{result in} performance degradation when \rev{categorizing} images from underrepresented regions \cite{denton2020bringing}.
    \item \textit{Measurement biases} can occur when choosing, collecting, and calculating features and labels for the prediction problem. For example, in medical \rev{contexts}, diagnosis  is \rev{frequently a stand-in for} a health condition; yet, \rev{some} gender and \rev{race} groups \rev{face elevated risks} of misdiagnosis, or underdiagnosis \cite{hoffman2016racial}.
    \item \textit{Aggregation biases} can occur when a \rev{universal} model is \rev{applied to that should be differentiated based on underlying user groups.} For example, \rev{when training} natural language processing models \rev{on} generic data, \rev{the nuances and contextual meanings of street slang can be lost} \cite{frey2020artificial}.
    \item \textit{Learning biases} can occur when modeling choices amplify performance \rev{gaps} across user segments. For example, \rev{prioritizing privacy in a model can diminish the impact of} underrepresented groups data \cite{bagdasaryan2019differential}.
    \item  \textit{Evaluation biases} can occur when the benchmark population is not representative of the \rev{target} population. For example, dark-skinned women comprise only a small percentage of popular facial image benchmarks, leading to worse \rev{intersectional} performance of commercial facial analysis tools \cite{buolamwini2018gender}.
    \item \textit{Deployment biases} can occur when there exists a mismatch between the problem a model is designed to solve and how it is actually utilized. For example, risk assessment tools in criminal justice can be used in ``off-label'' ways, such as determining the length of a sentence \cite{collins2018punishing}.
\end{itemize} 
  
In the following section, we introduce the use case through which we explore bias in PI for mHealth. We then show empirically and analytically how Suresh and Guttag's seven sources of bias translate in the PI domain.

\subsection{Exploring Bias through the Largest Digital Biomarkers mHealth Dataset\label{useCase}}
\rev{To enable our analysis} of bias in the PI life cycle, we require an indicative use case. For this purpose, we utilize the MyHeart Counts dataset \cite{hershman2019physical}, the largest collection of digital biomarkers in the mHealth domain to date, enabling us to perform the most comprehensive analysis of bias across diverse user demographics. Nevertheless, \mrev{our methodology can be potentially generalized to other PI datasets} (\S\ref{generalizability}). 

\subsubsection*{\textbf{Data Description}} Up till recently, \rev{generic}, population-scale PI datasets were \rev{uncommon} due to \rev{cost}, privacy concerns, and data protection regulations. \rev{Existing} open datasets \rev{were small- to medium-sized} \cite{vaizman2018extrasensory,wang2014studentlife} or domain-constrained, \rev{e.g.,} to human-activity recognition (HAR) \cite{anguita2013public}. \rev{The release} of data from the MyHeart Counts Cardiovascular Health Study from 50K participants in the \rev{US}, \rev{changed this situation}. Participants completed surveys and a 6-minute walk test and contributed PI data via a \rev{mobile} application. 
Approximately 1 out of 10 participants ($N=4920$) shared their basic \rev{PI} data\rev{, such as} step count, distance covered, burned calories, and flights climbed. We combine these data with survey responses to attain the following user attributes: gender, ethnicity, age, BMI, and health conditions, such as heart condition, hypertension, joint problem, and diabetes. 

\begin{table}[t]
\caption{The protected attributes in the MyHeart Counts data. For the purpose of the bias analysis, we binarize non-binary attributes to ensure a sufficient sample size per group and compatibility with popular bias metrics.}
\label{tab:protectedAttributes}
\resizebox{\textwidth}{!}{%
\begin{tabular}{llll}
 & \multicolumn{1}{c}{\textit{Original Protected Attribute Values}} & \multicolumn{2}{c}{\textit{Binarized Protected Attribute Values}} \\ \hline
\multicolumn{1}{l}{\textbf{Attribute}} & \textbf{Original Groups} & \multicolumn{1}{l}{\textbf{Majority Group}} & \textbf{Minority Group} \\ \hline
\multicolumn{1}{l}{Gender} & Male, Female, N/A & \multicolumn{1}{l}{Male} & Female \\ \hline
\multicolumn{1}{l}{Ethnicity} & White, Asian, Black, Hispanic, American Indian, Pacific Islander, Other, N/A & \multicolumn{1}{l}{White} & Non-white \\ \hline
\multicolumn{1}{l}{Age} & Integer Number, N/A & \multicolumn{1}{l}{<65}  (lower risk of complications) & >=65 (higher risk of complications) \\ \hline
\multicolumn{1}{l}{BMI} & Real Number (height and weight), N/A & \multicolumn{1}{l}{<18.5 or =>25 (unhealthy)} & =>18.5 and <25 (healthy) \\ \hline
\multicolumn{1}{l}{Heart Condition} & Yes, No, N/A & \multicolumn{1}{l}{No} & Yes \\ \hline
\multicolumn{1}{l}{Hypertension} & Yes, No, N/A & \multicolumn{1}{l}{No} & Yes \\ \hline
\multicolumn{1}{l}{Joint Problem} & Yes, No, N/A & \multicolumn{1}{l}{No} & Yes \\ \hline
\multicolumn{1}{l}{Diabetes} & Yes, No, N/A & \multicolumn{1}{l}{No} & Yes \\ \hline
\end{tabular}%
}
\end{table}

\subsubsection*{\textbf{Data Preprocessing}} To ensure sufficient sample size per user group and compatibility with bias metrics, we \rev{binarize} non-binary user attributes, such as ethnicity, age, and BMI, as seen in Table~\ref{tab:protectedAttributes}. This grouping creates two user groups per protected attribute, namely a majority group (``privileged'') and a minority group (``unprivileged''). Note that the usage of the term ``privilege'' in this work does not necessarily coincide with real-world ``privilege''. For example, users with unhealthy BMI are the majority, \rev{and hence ``privileged''} user segment in our dataset, whereas one could argue that the opposite applies in reality.

\begin{table}[t]
\caption{Example input data for the physical activity prediction use case. The step counts per hour for the past 48 hours are the features, and the total number of the next day's steps is the label. The user ID and timestamps are not used \rev{for} learning.}
\label{tab:inputData}
\resizebox{.5\textwidth}{!}{%
\begin{tabular}{llllll}
\hline
\multicolumn{5}{c}{\textbf{Features}} & \multicolumn{1}{c}{\textbf{Label}} \\ \hline
\multicolumn{1}{l}{user\_{id}} & \multicolumn{1}{l}{timestamp} & \multicolumn{1}{l}{steps at t-48h} & \multicolumn{1}{l}{...} & steps at t-1h & next day's steps \\ \hline
\multicolumn{1}{l}{1} & \multicolumn{1}{l}{23-11-2022} & \multicolumn{1}{l}{1040} & \multicolumn{1}{l}{...} & 300 & 8500 \\ \hline
\end{tabular}%
}
\end{table}

\subsubsection*{\textbf{Data Labeling}} The MyHeart Counts dataset does not introduce any prediction \rev{tasks}. To this end, we select the \textit{next-day physical activity prediction from historical data} use case \cite{bampakis2022ubiwear,vasdekis2022wemod} for model training. In other words, based on the user's past activity, we try to predict how many steps they will perform the next day (see Table~\ref{tab:inputData}), \rev{e.g., to enable personalized goal-setting}. Basic digital behavioral biomarkers, \rev{such as steps,} are easy to collect and commonplace in the literature, enabling the reproducibility of our findings. Also, \rev{they} are the largest available sensed modality in the My Heart Counts dataset, allowing us to \rev{exploit more} data for \rev{our} analysis. \mrev{At the same time, according to the World Health Organization (WHO), physical activity has significant health benefits for hearts, bodies, and minds, contributing to preventing and managing noncommunicable diseases such as cardiovascular diseases, cancer, and diabetes \cite{who2012report}.
Strikingly, physical inactivity has been identified by the WHO as the fourth leading risk factor for global mortality, accounting for 6\% of deaths globally \cite{world2010global}, highlighting the importance of the selected use case.}

\subsubsection*{\textbf{\rev{Bias Measures}}} \rev{}
\rev{To measure bias, ML researchers have quantified fairness metrics that operationalize fairness definitions (See Appendix~\ref{ap:definitions}). For this work, we utilize the widely used \rev{\textit{Disparate Impact Ratio (DIR)}}, which is the ratio of base or selection rates between unprivileged and privileged groups, assuming equal ability to perform physical activity across demographics:}
\begin{equation*}
\text{Disparate Impact Ratio}=\frac{\Pr(y^+\mid G0)}{\Pr(y^+\mid G1)}
\end{equation*}
where $y^+$ is the actual or predicted positive outcome label (base or selection rate, respectively), $G0$ is the minority (protected) group, and $G1$ is the majority group. Values \rev{lower} than 1 \rev{mean} the majority group has a higher proportion of positive outcomes; a value of 1 indicates demographic parity. 
For example, a value of 0.8 for a dataset with gender as protected attribute means that for every male receiving a high activity goal, only 0.8 females do so. According to the ``4/5 rule'' \cite{castelnovo2022clarification}, accepted values \rev{lie} within [0.8,1.25], but such ranges are not universally accepted and \rev{are context-dependent} \cite{equal1990uniform}. 

\section{Exploring Bias in Personal Informatics Data Generation}\label{dataBias}

Bias in the data generation stream can take the form of historical, representation, and measurement biases, as seen in Figure~\ref{fig:biasData}. In this section, we explore all three sources, \rev{answering to} \rev{RQ1:} \textit{Are PI data susceptible to biases?}
\subsection{Historical Bias\label{historicalBias}}
\rev{Historical biases are domain- rather dataset-dependent, and hence not necessarily quantifiable}. Hence, for completeness, we state the main findings of the related literature \rev{on the PI domain.}

\subsubsection*{\textbf{Physical Activity Inequalities}} Physical activity data, such as step counts, are among the most common digital biomarkers, \rev{and} constitute the majority of the extracted \rev{MyHeart Counts data, with 4920 users of step tracking compared to 626 users of sleep tracking.} However, inequalities in physical activity are \rev{well-documented in literature} \cite{guthold2018worldwide,althoff2017large,world2019global}. \citet{althoff2017large} reveal variability in physical activity worldwide, where reduced activity in females explains a large portion of the observed activity inequality. Overall, the World Health Organization reports that ``girls, women, older adults, people of low socioeconomic position, people with disabilities and chronic diseases, marginalized populations, indigenous people and the inhabitants of rural communities often have less access to safe, accessible, affordable and appropriate spaces and places in which to be physically active'' \cite{world2019global}. Such \rev{real-world} inequalities can \rev{manifest} into the behavioral data we \rev{use to train our models.} 

\subsubsection*{\textbf{The Digital Divide}} Similarly, as the world rapidly digitalizes, it threatens to exclude those that remain offline. Almost half the world’s population, the majority of them women or citizens of developing countries, are still disconnected \cite{mohammed2021almost}. Even in the connected world, male internet users outnumber their female counterparts. This ``digital divide'' encompasses even more discrepancies, such as the digital infrastructure quality and connectivity speed in rural or remote areas and the required skills to navigate technology \cite{chetty2018bridging}. \rev{Therefore, it is clear that} technological systems, including PI, \rev{are limited in their ability to capture the diversity of the world population,} due to pre-existing inequalities in digital access and literacy. 

\subsubsection*{\textbf{BYOD Study Design Biases}} PI technologies are \rev{used} for data collection in clinical research, resulting in newfound demographic imbalances. Studies adopting a bring-your-own-device (BYOD) design, such as MyHeart Counts, are more user-friendly, achieve better participant compliance, potentially reduce the bias of introducing new technologies, and accelerate data collection from larger cohorts \cite{demanuele2022considerations,cho2022demographic}. However, \citet{cho2022demographic} identifies significant demographic disparities regarding race (50-85\% white cohorts) in BYOD studies. Their findings align with the reported demographic divide existent in the composition of wearable users. Even though the gap is narrowing, a \rev{recent} report \cite{ericsson2016} documents that \rev{most} existing wearable users are fit adults between 25–34 and that whilst females are more likely to own activity trackers, 63\% of smartwatch owners are male.  Hence, the technology and participant cohorts in PI \rev{in the context of BYOD} studies subject datasets to the same biases exposed in the activity inequality and the digital divide literature.

\subsection{Representation Bias\label{representationBias}}

We discuss representation biases across three dimensions: misrepresented, underrepresented, and unevenly sampled populations. 
\subsubsection*{\textbf{Misrepresented Populations}} 
\begin{figure}[t]
    \centering
    \begin{minipage}{0.5\textwidth}
        \centering
        \includegraphics[width=.95\textwidth]{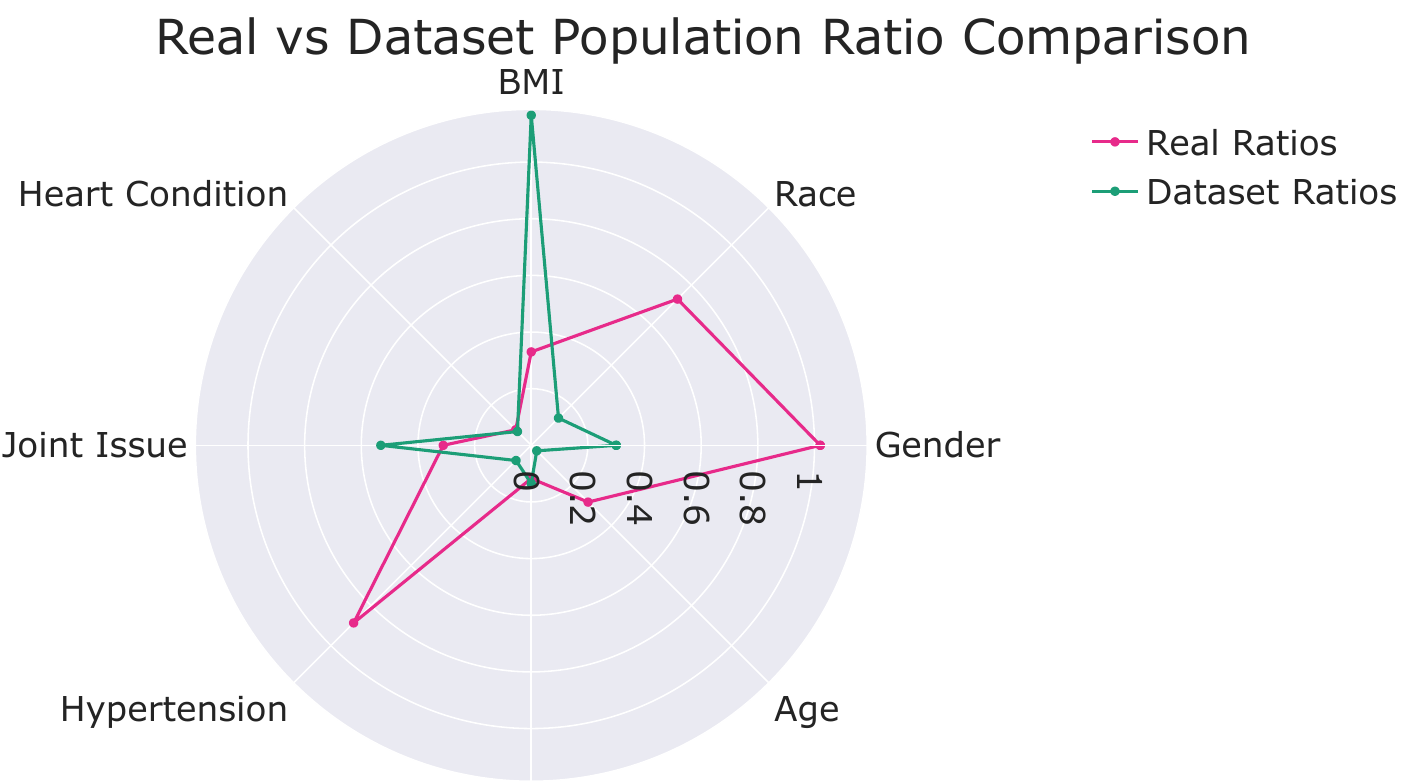} %
        \caption{Real (pink) versus dataset (green) ratios \rev{across} population segments in the MyHeart Counts dataset. The ratio is calculated as the number of the minority class divided by the number of the majority class \rev{samples}. \rev{Larger distances indicate larger deviations. Gender, age, race, and to a lesser extent hypertension and BMI show representation bias.}\label{fig:representation1}}
    \end{minipage}\hfill
    \begin{minipage}{0.4\textwidth}
        \centering
        \includegraphics[width=.95\textwidth]{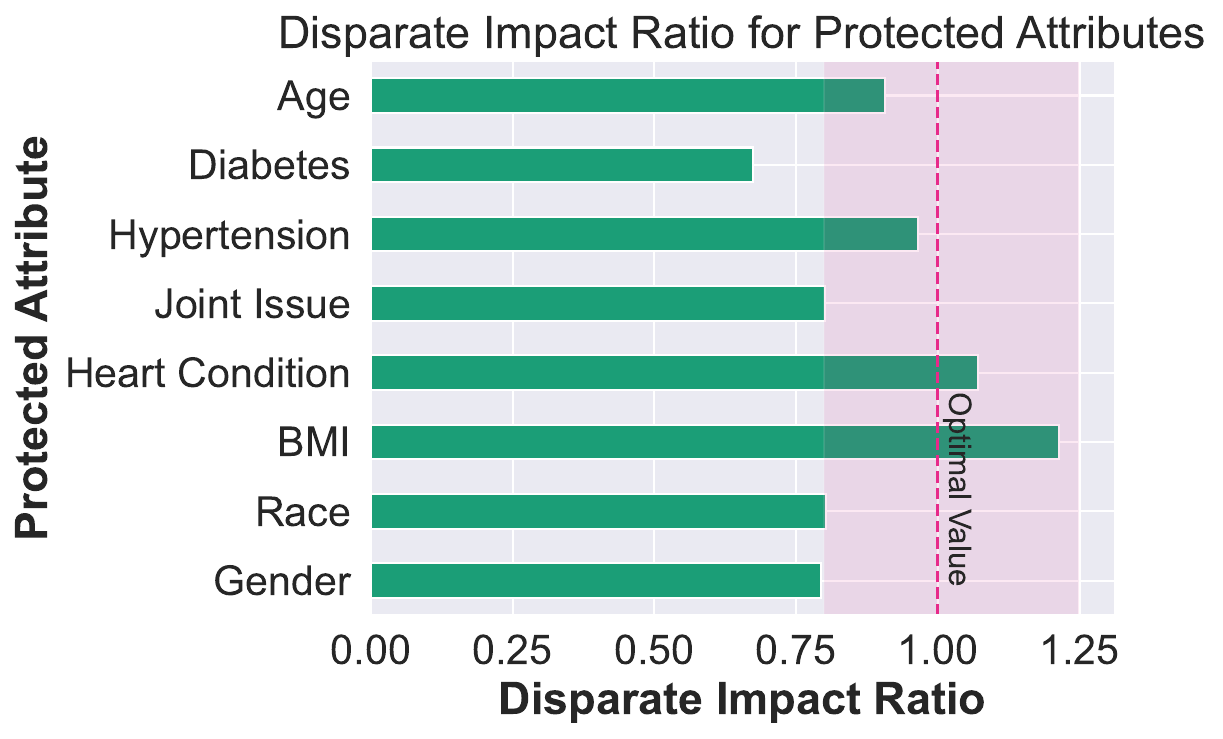} %
        \caption{A bar plot showing the DIR (ratio of base rates) per protected attribute. We notice that there exist biases in columns for diabetes patients, users with joint issues, and non-white minorities. While the data is borderline biased against women and people with unhealthy weight (underweight, overweight, or obese).\label{fig:representation3}}
    \end{minipage}
\end{figure}

Representation bias can emerge when the sample population does not reflect the general population (bias in rows). In the MyHeart Counts dataset, we compare the ratios of majority and minority user segments as defined in Table~\ref{tab:protectedAttributes} with the real-world ratios extracted from \rev{US} population censuses. Specifically, we utilize the \rev{US} Census Bureau (gender, race, and age \cite{bureauGender} distributions), the Centers for Disease Control and Prevention (BMI \cite{fryar2020prevalenceBMI,fryar2020prevalenceUnderweightBMI}, joint issues \cite{theis2021prevalenceJOINT}, hypertension \cite{centers2019hypertension}, and diabetes \cite{us2020nationaldiabetes}), and the American Heart Association (heart condition \cite{tsao2022heart}) data. Figure~\ref{fig:representation1} showcases the results of this comparison in a radar plot. For example, while in the general \rev{US} population, we have approximately 1 female per 1 male (ratio of 1.0 in pink), in the MyHeart Counts data, we have 0.2 females per 1 male, highlighting the \rev{substantial} underrepresentation of women. The same applies to race, age, and hypertension segments, where the minority classes in the dataset (non-white users, users less than 45, and users with hypertension, respectively) do not reflect real-world ratios. An interesting finding is that, while in the \rev{US}, there exist approximately 0.3 underweight, overweight, or obese people for every person with normal weight, in the dataset, this ratio is doubled. Hence, \rev{possibly} due to historical biases and design choices, our analysis of the MyHeart Counts data (Figure~\ref{fig:representation1}) provides evidence that PI datasets might not \rev{be representative of} target populations.

\subsubsection*{\textbf{Underrepresented Populations}} 
\begin{figure}[t]
  \centering
  \includegraphics[width=.9\linewidth]{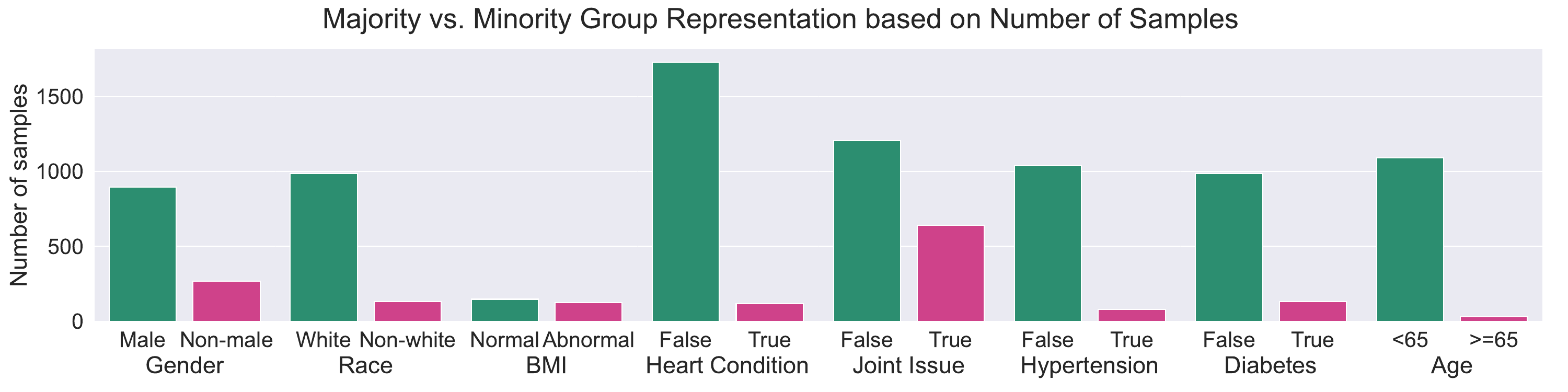}
  \caption{A bar plot showcasing the number of samples per user segment split based on various protected attributes. We see significant underrepresentation of minority user segments across almost all attributes.\label{fig:representation2}}
\end{figure}

PI datasets can still include underrepresented groups (bias in rows) even if sampled perfectly. Figure~\ref{fig:representation2} shows significant imbalances in the number of samples between minority and majority user segments across almost all protected attributes. We notice that even for representative sampling, e.g., users with joint or heart problems, the minority group is still significantly underrepresented. Thus, the model \rev{might} be less robust for those users because it has \rev{less} data to learn from. Overall, we see that the MyHeart Counts data are skewed towards \textit{white, fit males}, which needs to be considered in the model-building phases. \rev{An ideal PI dataset should be representative of the target population while having enough minority samples. However, building large-scale PI datasets is challenging due to the required effort and cost.}

\subsubsection*{\textbf{Unevenly Sampled Populations}} Even if sampling is representative and equal (e.g., 50\% male and 50\% female users), the dataset can still suffer from representation bias if the sampling method is uneven, \rev{e.g., active males but inactive females} (bias in columns). This is also the case in MyHeart Counts. Figure~\ref{fig:representation3} shows the DIR value per protected attribute (i.e., the ratio of recorded high \rev{activity} for unprivileged versus privileged groups). For our use case, a value of $DIR<0.8$ means that the \rev{minority} sample is significantly less active than \rev{the} majority \rev{sample}. \rev{Specifically}, in the MyHeart Counts data, diabetes patients, users with joint issues, racial minorities, and to a smaller extent, women, racial minorities, and overweight and obese users systematically perform lower step counts in the dataset compared to their majority counterparts. On the contrary, users of different age groups with or without hypertension or heart issues do not differ significantly \rev{regarding} step counts in the data.

\subsection{Measurement Bias\label{measurementBias}}

\rev{In this section,} we focus on the input modalities and their accuracy and discrepancies during data collection. 

\subsubsection*{\textbf{Device Differences}} Data \rev{in the MyHeart Counts HealthKit dataset} originate from different sources \rev{(i.e., 33\% iPhone, 11\% Apple Watch, and 56\% third parties}. 
\rev{iPhones use integrated sensors, including} accelerometer, gyroscope, GPS, and magnetometer to detect and calculate step counts. \rev{The motion coprocessor unit reads the sensor data and communicates with the CMMotionActivityManager to classify user activity.} This process cannot be fully replicated in Apple watches due to differences in placement, fit, and usage habits. \rev{Phones may} underestimate step counts due to non-carrying time, while watches \rev{have been found} more accurate for measuring daily step counts for healthy adults \cite{amagasa2019well,duncan2018walk,veerabhadrappa2018tracking}. MyHeart Counts HealthKit data \rev{show} statistically significant differences ($p<0.05$) in watch ownership \rev{across segments} based on gender (46\% of males have at least one watch entry \rev{vs.} 28\% \rev{females}), heart condition (38\% with \rev{vs.} 26\% without), and ethnicity (41\% non-white \rev{vs.} 36\% white).

\subsubsection*{\textbf{Model Differences}} 
\begin{figure*}[t!]
    \centering
    \begin{subfigure}[t]{0.45\textwidth}
        \centering
        \includegraphics[width=.65\linewidth]{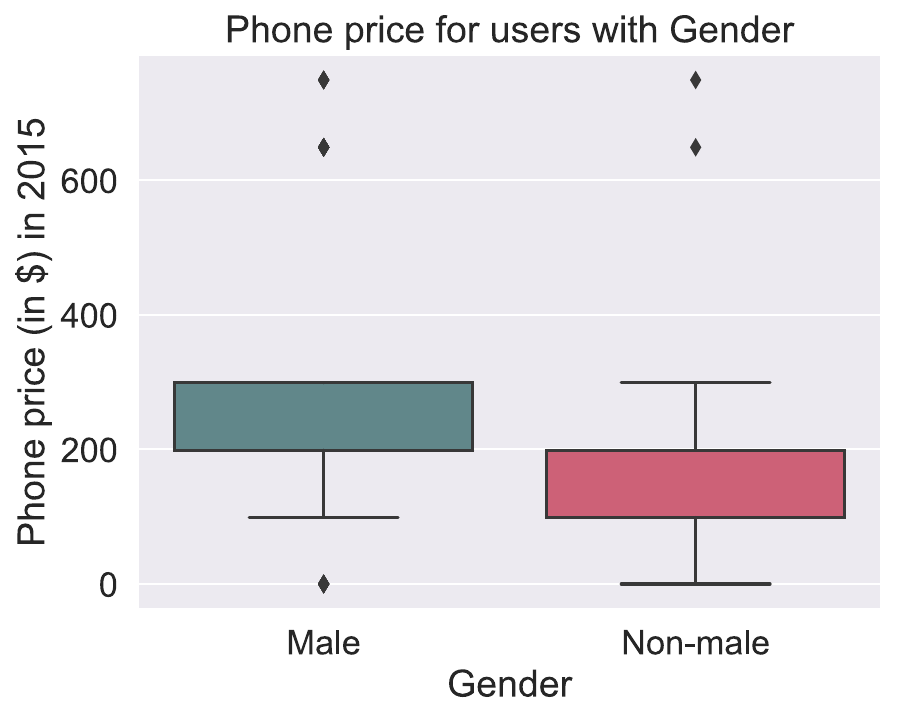}
    \end{subfigure}%
    ~ 
    \begin{subfigure}[t]{0.45\textwidth}
        \centering
        \includegraphics[width=.65\linewidth]{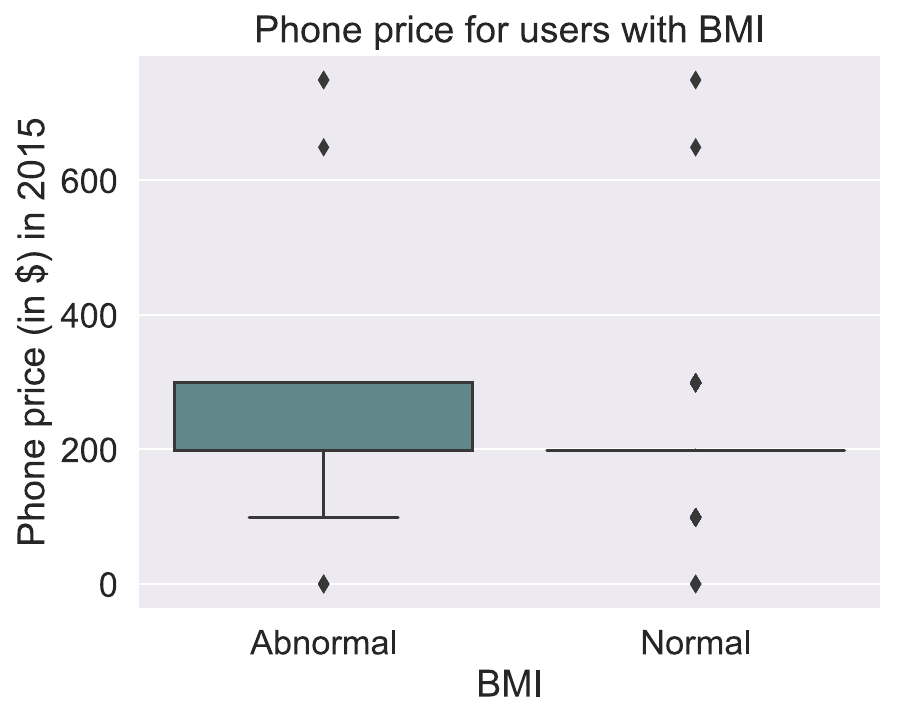}
    \end{subfigure}
    \caption{Differences in the price of participants' phones as of September 2016 based on gender (left) and BMI (right). Females and people with BMI within the normal range tend to own older and cheaper phones with fewer capabilities.\label{fig:boxplotGenderBMI}}
\end{figure*}

Accuracy differences have been reported across consecutive generations of phone devices \cite{duncan2018walk}. Incremental hardware changes may increase the quantity, modality, and quality of data available for the device to \rev{classify user activity}. \rev{For instance, the existence of specialized coprocessors, ``always-on'' capabilities, and revised recognition algorithms in newer phones can improve classification accuracy.} In the MyHeartCounts data, we encounter various phone models, spanning \rev{five generations}. We identify statistically significant differences ($p<0.05$) \rev{in phone ownership} based on gender and BMI. Specifically, females and people with normal BMI tend to own older and cheaper phones with fewer capabilities (see Figure~\ref{fig:boxplotGenderBMI}).

\subsubsection*{\textbf{General Input Modality Differences}} Finally, most MyHeart Counts data comes from third parties, such as alternative wearables or fitness and well-being apps. This is common in the PI domain, given the abundance and heterogeneity of available data sources.
\rev{In our use case,} we identify statistically significant differences \rev{in third party usage} based on gender (91\% \rev{males} have at least one third party entry compared to 85\% \rev{females}) and diabetes condition (97\% with \rev{vs.} 90\% without). However, different input devices or apps are proven to have different accuracies, likely to create measurement accuracy discrepancies between different users \cite{el2015currently}.

\section{Exploring Bias in Personal Informatics Model Building and Implementation\label{modelBiases}}
Bias in the model building and implementation stream can take the form of aggregation, learning, evaluation, and deployment biases, as seen in Figure~\ref{fig:biasModel}. In this section, we discuss all four sources, providing an answer to RQ3, namely: \textit{Do ML models inherit PI data biases? Do they mitigate, propagate, or maybe even amplify them?}
\subsection{Aggregation Bias}\label{aggregation}

We evaluate aggregation bias by plotting the DIR (selection rate, i.e., rate of high activity goals predictions) for different user segments' predictions based on heart conditions, hypertension, joint issues, diabetes, race, BMI, gender, and age. We utilize two baseline models to capture the notions of ``fairness through awareness'' \cite{dwork2012fairness} and ``fairness through unawareness'' \cite{kusner2017counterfactual}. In fairness through awareness, fairness is captured by the principle that similar individuals should have similar classification outcomes. In our use case, the similarity is defined based on user demographics in the absence of other features. In practical terms, the aware model is trained on a feature set that includes protected attributes per user. On the other hand, fairness through unawareness is satisfied if no \rev{protected} attributes are explicitly used in the learning process \cite{verma2018fairness}.

\subsubsection*{\textbf{Models' Description}} Our baseline models \rev{and hyperparameters} are sourced from prior work in the field of physical activity prediction that benchmarked six distinct learning paradigms from traditional ML models to advanced deep learning architectures \rev{on the MyHeart Counts dataset \cite{bampakis2022ubiwear}}. \rev{For the scope of our work, we choose their best-performing model}, a Long Short-Term Memory (LSTM) recurrent neural network, \rev{that} achieves a Mean Absolute Error (MAE) of 1087 steps, beating previous state-of-the-art approaches by 67\% on the task of physical activity prediction. \rev{For more details on the model architecture see Appendix~\ref{ap:modelArchitecture}.}

\begin{figure}[t]
    \begin{minipage}{0.48\textwidth}
        \centering
          \includegraphics[page=1, width=.9\linewidth]{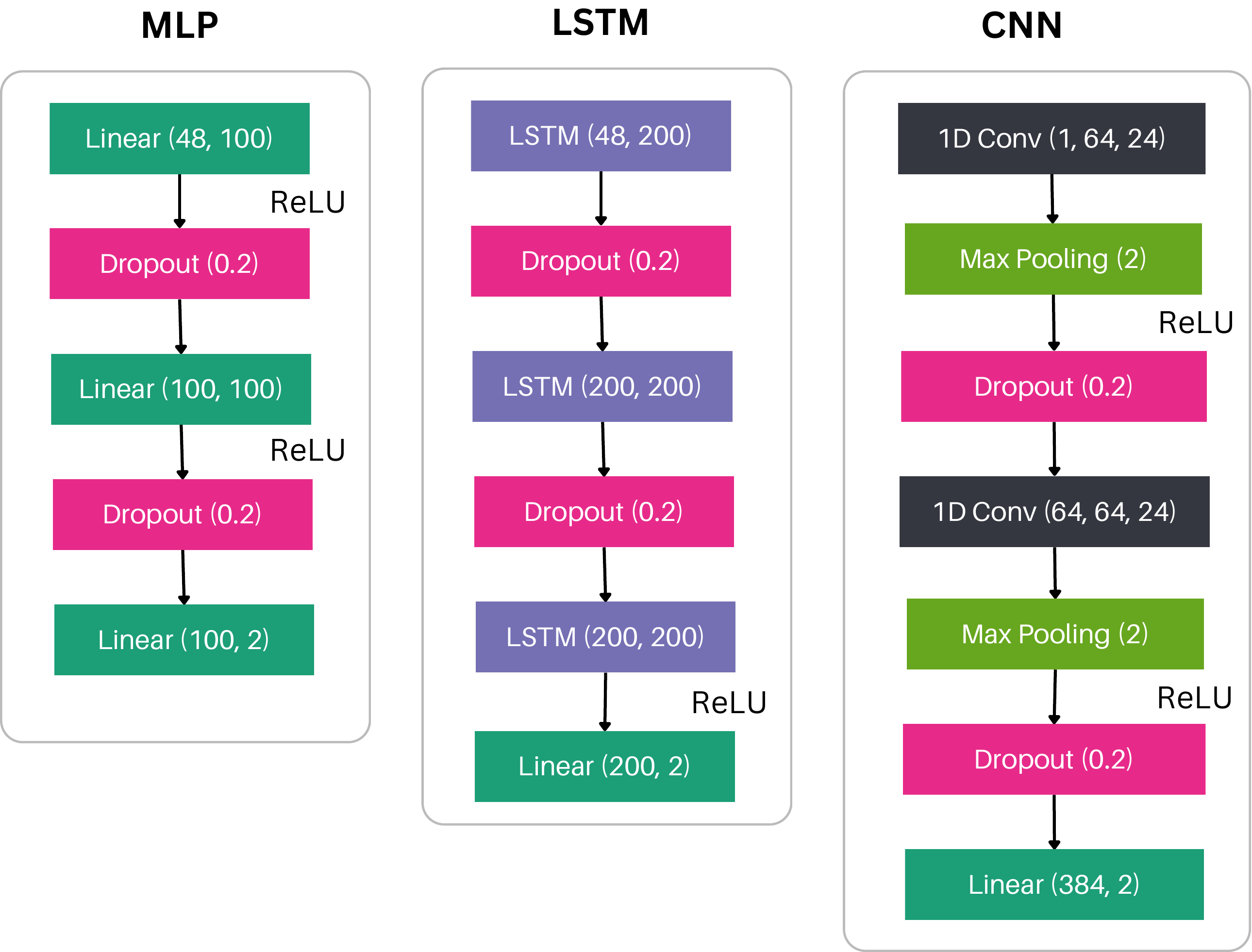}
          \caption{\rev{Alternative model architectures evaluated: MLP, LSTM, and CNN.}\label{fig:modelsAlt}}
    \end{minipage}\hfill
    \begin{minipage}{0.48\textwidth}
          \centering
          \includegraphics[width=.82\linewidth]{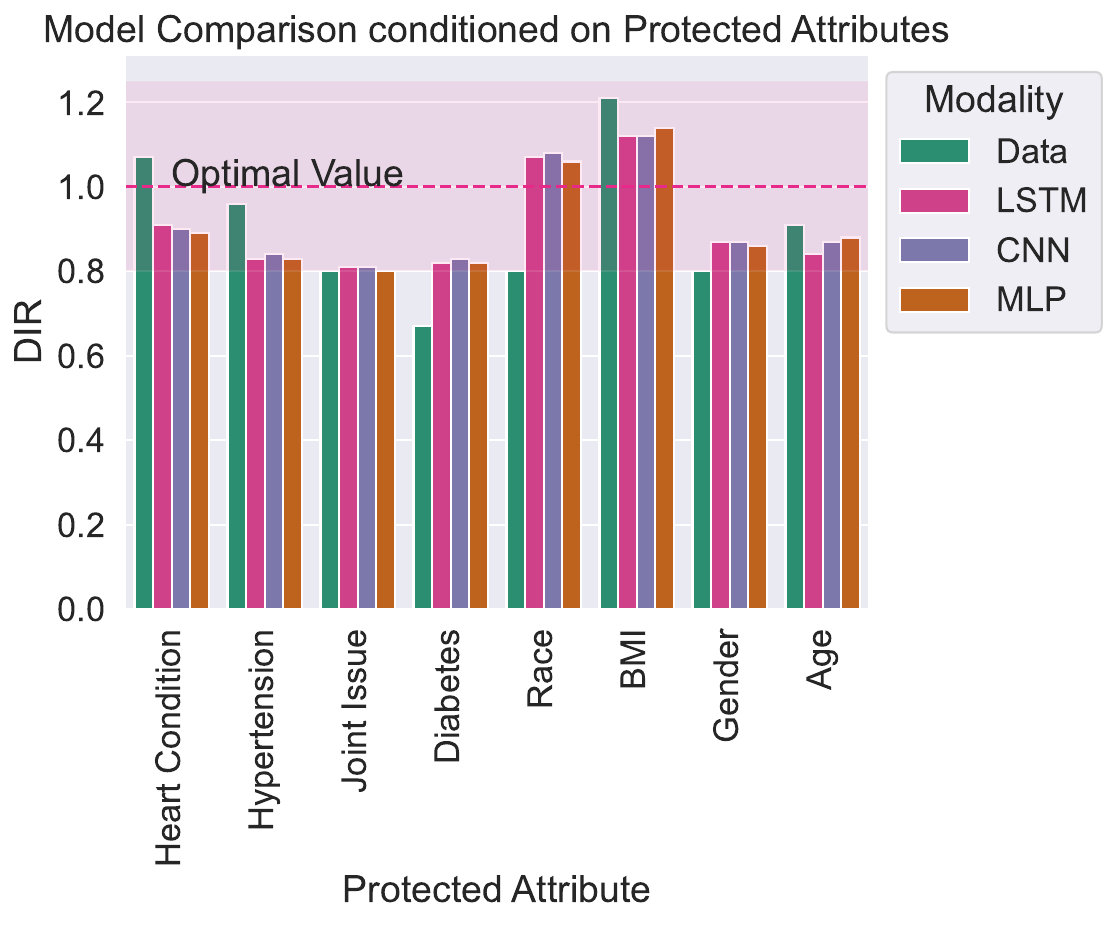}
          \caption{\rev{A comparison of DIR between data and models shows no significant differences in terms of fairness (DIR).}\label{fig:models}}
    \end{minipage}
\end{figure}
\rev{To further validate our model choice (LSTM), we also conduct comparative fairness assessments between three deep learning models (Figure~\ref{fig:modelsAlt}):}
\rev{
\begin{itemize}
    \item 3-layered Multilayer perceptron (MLP) with dropout
    \item 2-layered Convolutional Neural Network (CNN) with max-pooling and dropout
    \item 3-layered Long Short-Term Memory (LSTM) with dropout
\end{itemize}
}
\rev{As per Figure~\ref{fig:models}, both alternatives perform similarly to the ``unaware'' LSTM model regarding fairness metrics (i.e., only 0-2\% deviation in DIR), indicating the generalizability of our claims across learning paradigms.}

\subsubsection*{\textbf{Single Attribute Biases}}  
\begin{figure}[t]
    \centering
    \begin{minipage}{0.48\textwidth}
        \centering
        \includegraphics[width=\textwidth]{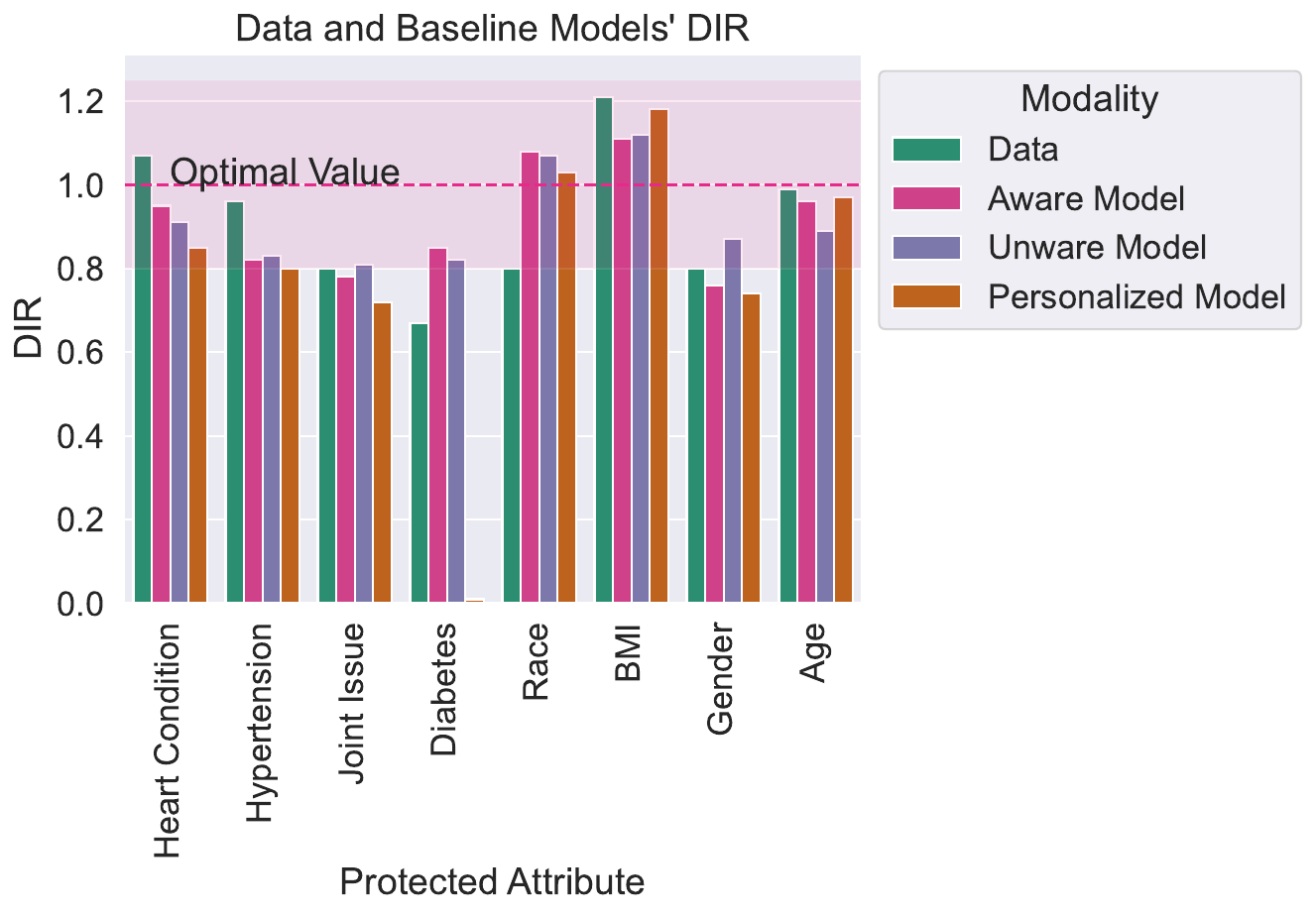} %
  \caption{\rev{DIR} comparison between data, \rev{aware, unaware baseline, and personalized} models. We see that the ``one-size-fits-all'' models propagate or, in some cases, amplify existing representation biases.\label{fig:DIRbaseline}}
    \end{minipage}\hfill
    \begin{minipage}{0.48\textwidth}
        \centering
        \includegraphics[width=1.1\textwidth]{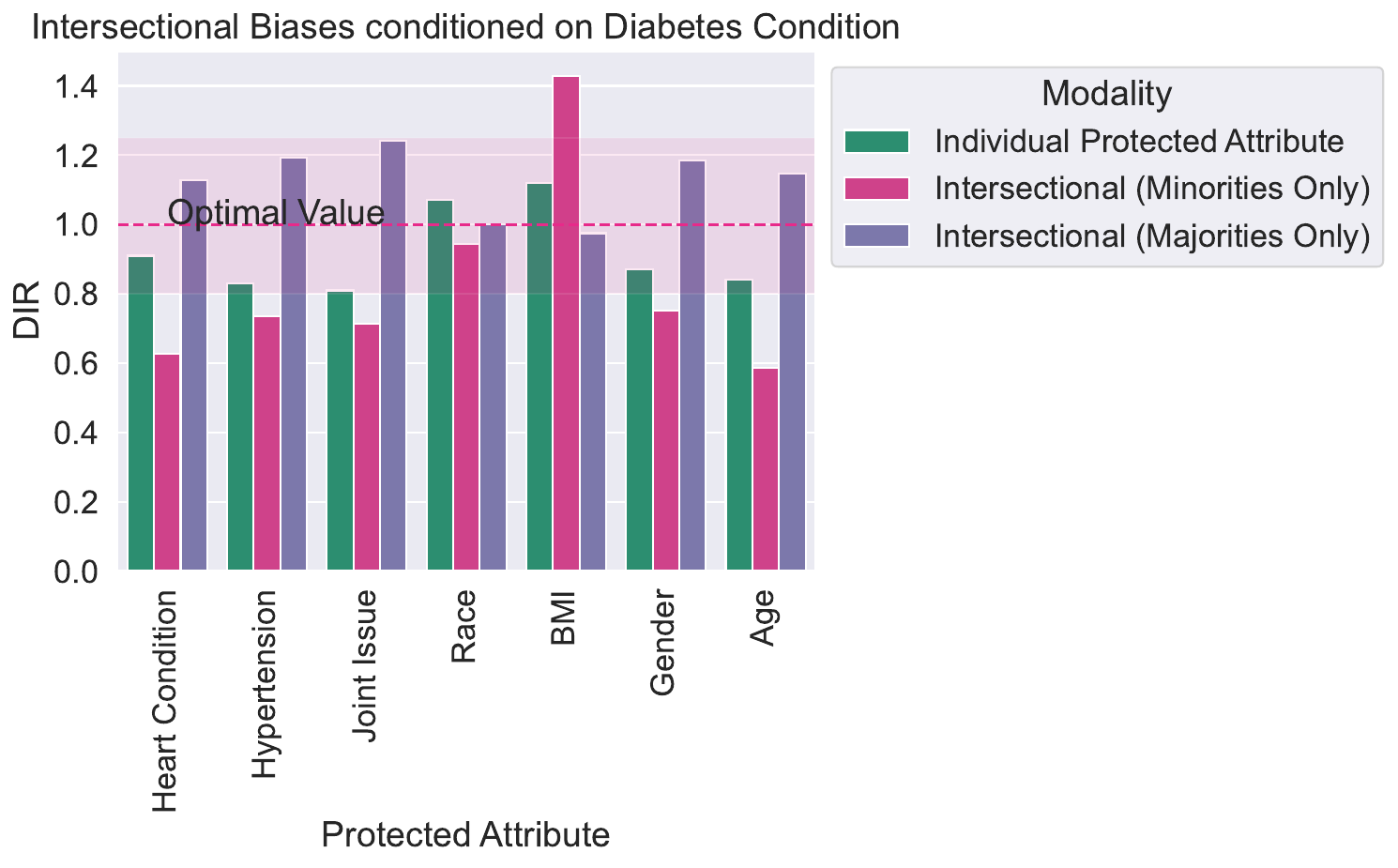} %
  \caption{\rev{DIR unaware baseline model} comparison between \rev{single-attribute and intersectional} user groups. Intersectional groups are drawn from the minority and majority classes. The ``one-size-fits-all'' models' biases are more prevalent in intersectional \rev{models}.\label{fig:DIRbaselineIntersectional}}
    \end{minipage}
\end{figure}

\rev{Figure~\ref{fig:DIRbaseline} presents experimentation} results concerning ML model biases measured via DIR. \rev{Specifically,} aware learning models are not foolproof against data biases in most cases (joint issues, diabetes, gender), and even amplify them for \rev{specific} protected attributes (hypertension); (2) Even excluding protected attributes from the training process of unaware models does not guarantee unbiased results in line with prior work \cite{pedreshi2008discrimination}. Fairness through unawareness is also ineffective due to the presence of proxy features, \rev{i.e., features} that work like proxies for protected attributes. Through \rev{them}, bias propagates from data to models: for example, a person's walking behavior (measured in step counts) is a good predictor of gender, BMI, and age, which can thus be inferred, despite being hidden during training~\cite{thomas2022fitbit}. Overall, diabetes patients have the largest bias gap compared to \rev{non-diabetic users}, partially attributed to their highly biased training data. Yet, users with hypertension have the largest difference between data and model biases since models trained on seemingly unbiased \rev{data} introduce bias during the learning process.

\subsubsection*{\textbf{Intersectional Biases}} We also examine intersectional biases, as shown in Figure~\ref{fig:DIRbaselineIntersectional}; namely we quantify the biases of the unaware model \rev{conditioned on} protected attribute combinations. Specifically, we consider two attributes at a time and two different combination strategies: \textit{minority-minority vs. rest} (e.g., diabetic women) and \textit{majority-majority vs. rest} (e.g., non-diabetic men). Keeping the diabetes attribute fixed, \rev{our results} highlight the widening intersectional biases for \rev{individuals} who belong to \rev{multiple} minorities (in pink) across almost all attributes (with \rev{the} exception of BMI, where \rev{individuals} with unhealthy BMI are the majority group, despite usually being considered unprivileged in practice). The largest gap appears in \rev{individuals} with more than one health condition, such as diabetic heart patients and diabetic patients aged 65+. At the same time, \rev{individuals} who do not belong to any minority groups (in purple), benefit across all attributes.
The trends in aggregation bias indicate that PI models do not tackle diverse user segments equally well and reflect or even amplify representation biases existing in the data, especially \rev{regarding} intersectional biases. 

\subsection{Learning Bias\label{learningBias}}
\rev{Personalization in prevalent in the PI literature}, straying from the ``one-size-fits-all'' mentality and its shortcomings, as discussed above. Contrary to generic models, personalized models are fine-tuned given the data of a single user or user segment. Accounting for such interindividual variability has been proven to dramatically improve prediction performance in various tasks within the PI domain, such as pain detection, engagement estimation, and stress prediction from ubiquitous devices data \cite{taylor2017personalized,shi2022toward,lopez2017multi}. Given the increasing popularity of the personalization paradigm, in this study, we investigate whether personalization as a modeling choice can amplify performance disparities across different user segments in the data, given the existence of representation bias.

\begin{figure}[t]
  \centering
  \includegraphics[width=.7\linewidth]{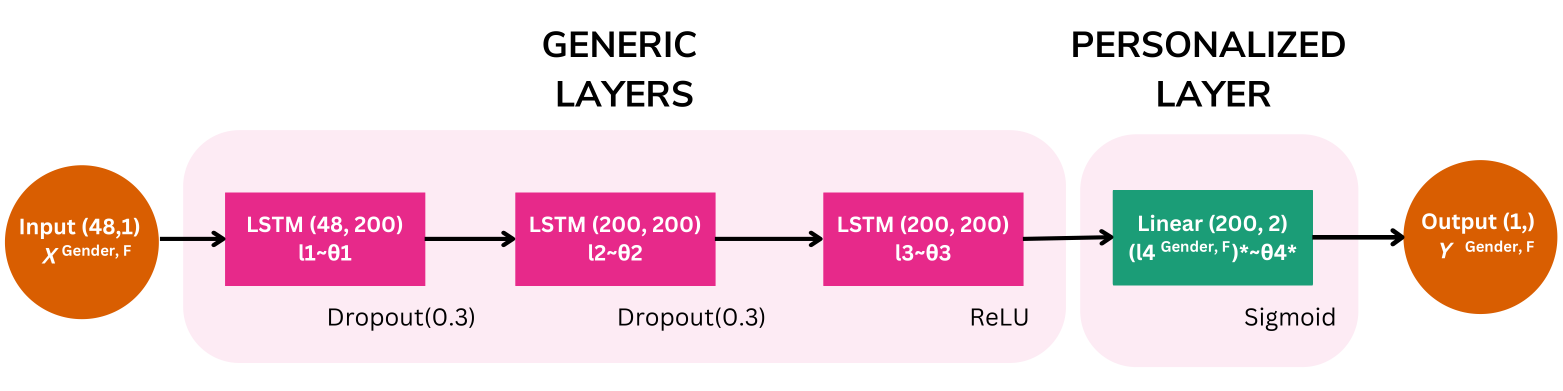}
  \caption{Our personalized deep learning architecture inspired by CultureNet \cite{rudovic2018culturenet}. \label{fig:personalization}}
\end{figure}
\subsubsection*{\textbf{Model Description}} We base our approach on the CultureNet package \cite{rudovic2018culturenet,rudovic2017measuring} for building generalized and culturalized deep models to estimate engagement levels from face images of children with Autism Spectrum Condition. Specifically, we utilize our deep LSTM model, which is trained on data from all users, we freeze the network parameters tuned to both minority and majority user groups, as described in Section~\ref{aggregation}, and then fine-tune the last layer, i.e., a linear fully-connected layer, to each user group separately based on the MyHeart Counts protected attributes (health condition, hypertension, joint issues, diabetes, race, BMI, gender, age). Figure~\ref{fig:personalization} delineates the personalization process. \rev{Appendix~\ref{ap:modelArchitecture} provides a formal definition of the learning process.}

\subsubsection*{\textbf{Single Attribute Biases}} While we could not identify significant performance benefits either for the privileged or unprivileged groups by utilizing personalization in our use case, we encountered significant bias shortcomings \rev{(Figure~\ref{fig:DIRbaseline})}. Specifically, across all protected attributes (with a borderline exception of race), personalized models \rev{are more biased than} either aware or unaware models or both. Users with diabetes \rev{present an extreme case}. The personalized model ``learns'' that \rev{diabetics are} less active than \rev{healthy users} in the dataset and thus provides \rev{solely} low activity goals, \rev{even to active diabetics}. The intuition behind this behavior \rev{lies in the training process}; personalized models amplify \rev{data representation biases} through fine-tuning. Our findings highlight that a common modeling choice in PI, such as personalization, can negatively affect biases and asks for bias-aware personalization approaches to rip the benefits of user tailoring without leading to biased results.

\subsection{Evaluation Bias\label{evaluationBias}}\label{sec:evalBias}
\subsubsection*{\textbf{Benchmark Selection}} \rev{ML models} are optimized on training \rev{and validation} data but evaluated on \rev{test} benchmarks \rev{\cite{deng2009imagenet,harper2015movielens}}. However, the ubiquitous computing community still suffers from a lack of \rev{larger} benchmarks \rev{beyond} HAR \cite{anguita2013public} and sleep classification \cite{zhang2018national,chen2015racial}. \rev{Also}, benchmarks within the community \rev{often do not represent} the target population. For example, within the fall detection domain, \rev{due to safety concerns,} datasets usually comprise imitated falls performed by younger people while they are deployed on older people \cite{sucerquia2017sisfall}. 

\begin{figure}[t]
  \centering
  \includegraphics[width=.45\linewidth]{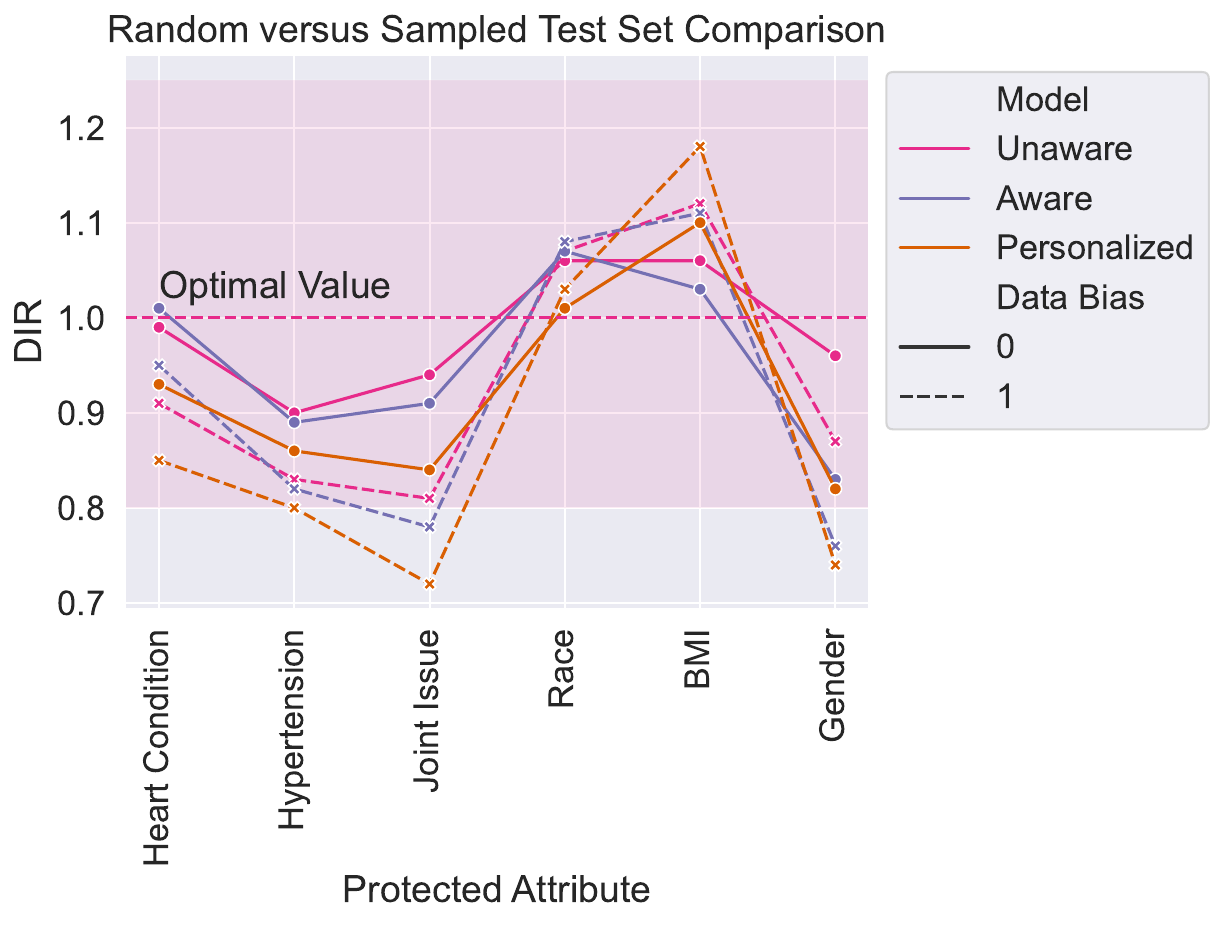}
  \vspace*{-0.1cm}
  \caption{A comparison of DIR between different test sets across models. \rev{``Ideal''} test sets in terms of data bias (\rev{dashed} lines) tend to hide imperfections in the trained models compared to the original, ``realistic'' test sets (\rev{continuous} lines).\label{fig:DIRtest}}
\end{figure}

Yet, a misrepresentative benchmark encourages \rev{deploying} models that perform well \rev{only on the} benchmark \rev{population}. To illustrate our point, given the lack of established benchmarks \rev{in PI}, we devise two distinct test sets for comparison purposes: our original, \rev{``realistic''} (random) test set, and a sampled subset of the latter (``ideal''), with demographic parity at base rate ($\text{DIR}=1.0$). We then evaluate our models on these two test sets. Figure~\ref{fig:DIRtest} presents the results of our experimentation, where it is clear that \rev{the ideal test set, imitating a} ``fair'' world, consistently shows \rev{lower bias than the ``realistic'' test set}. Better performance is defined as smaller deviations from the optimal $\text{DIR}$ value of $1.0$. Essentially, an ideal-world benchmark is ``hiding'' the imperfections of our trained model, which has been proven to propagate or even amplify biases based on the \rev{original, random test set}.

\subsubsection*{\textbf{Evaluation Metric Selection}} Evaluation bias can also emerge from the metric used to quantify the models' performance. For instance, group fairness hybrid metrics, such as error rates, are prone to imbalances and can hide disparities in other types of bias metrics, such as WAE metrics (see Appendix~\ref{ap:definitions}). Similarly, aggregate measures, such as accuracy, can hide subgroup under-performance or conceal shortcomings in \rev{other} metrics \cite{suresh2021framework}.

\subsection{Deployment Bias\label{deploymentBias}}

\subsubsection*{\textbf{Changing Deployment Scenarios}} \rev{PI's} most active research areas are Human-Activity Recognition and Sleep Classification. From this lens, \rev{False Positives (FP) and False Negatives (FN), i.e.,} Type I and Type II errors, respectively, in these scenarios are not critical, and models have been developed to maximize \rev{True Positives (TPs)}. This dominant view promotes deployment bias in novel use cases with the emergence of health-related intelligence embedded into PI systems. For example, given ECG sensor data and AFib detection functionality, Type II errors should be minimized to avoid loss of life.  It is thus critical to reassess the conceptualization of PI systems' evaluation practices and datasets and tailor them to their context.

\subsubsection*{\textbf{Development in Isolation}} \rev{ML} models for PI systems are built and evaluated as if they were fully autonomous. \rev{In} reality, they operate in a complex socio-ethical system moderated by institutions and human decision-makers, also known as the ``framing trap'' \cite{selbst2019fairness}. Users may share their mHeatlh data with physicians for interpretation and disease management. Despite good performance in isolation, they may lead to harmful consequences \rev{due to} human biases, \rev{e.g.,} confirmation bias. Specifically, physicians are more likely to believe AI that supports current practices and opinions \cite{parikh2019addressing}. At the same time, research shows that physicians’ perceptions about black male patients’ physical activity behavior were significant predictors of their recommendations for surgery, independent of clinical factors, appropriateness, payer, and physician characteristics \cite{van2006physicians}. Such complicated interconnections highlight how evaluating a system in isolation creates unrealistic notions of its benefits and harms.

\subsubsection*{\rev{\textbf{Biased Interpretation.}}}
\rev{Interpreting biased data can result in self-trackers making incorrect inferences or inappropriate tracking decisions. Discomfort with the information revealed and concerns about data quality ---which may not be consistent across demographics--- can lead to PI abandonment \cite{epstein2016beyond}. 
Additionally, discrepancies between users' expectations and biased data and subjectivity and uncertainty in data interpretation can fuel rumination (i.e., anxious self-attention and fear of failure), hindering self-improvement efforts and increasing the likelihood of abandonment. This is particularly relevant for health tracking and vulnerable populations, such as those with chronic illnesses, mental health conditions, and women facing fertility challenges, where the association of goals with identity and critical outcomes may increase the propensity for rumination~\cite{eikey2021beyond}.}

\section{Generalizability}
\label{generalizability}
This section aims to (i) demonstrate the straightforward applicability of our methodology to other datasets and (ii) reveal initial insights about the generality of our findings and future steps.

While our analysis was conducted on the MyHeart Counts dataset, \rev{some} of our findings can be \rev{potentially} generalized to other scenarios in PI and mHealth. To showcase this, we apply part of our experiments on two distinct datasets:
\begin{itemize}[leftmargin=*,nosep]
    \item \textbf{LifeSnaps} is a newly-released, \rev{medium-scale}, multi-modal dataset containing 71M rows of anthropological data, collected unobtrusively for the total course of more than four months by 71 participants. Based on data availability, we consider three protected attributes in Lifesnaps, namely gender, age, and BMI. Also, given the lack of official benchmark tasks, we consider the ``next-day physical activity prediction'' task for model training, same with the MyHeart Counts dataset.
    \item \textbf{MIMIC-III} is an established, large-scale clinical dataset consisting of information concerning more than 38K patients admitted to intensive care units (ICU) at a large tertiary care hospital in the \rev{US}. Based on data availability, in MIMIC-III, we consider six protected attributes, namely gender, ethnicity, language, insurance, religion, and age. Contrary to LifeSnaps or MyHeart Counts, there exists a public benchmark suite that includes four different clinical prediction tasks for MIMIC-III \cite{harutyunyan2019multitask}. For this analysis, we utilize the ``in-hospital mortality'' task as a binary classification equivalent to the ``next-day physical activity prediction'' task.
\end{itemize}

\begin{figure}[t]
\centering
\begin{subfigure}{.45\textwidth}
    \centering
    \includegraphics[width=\linewidth]{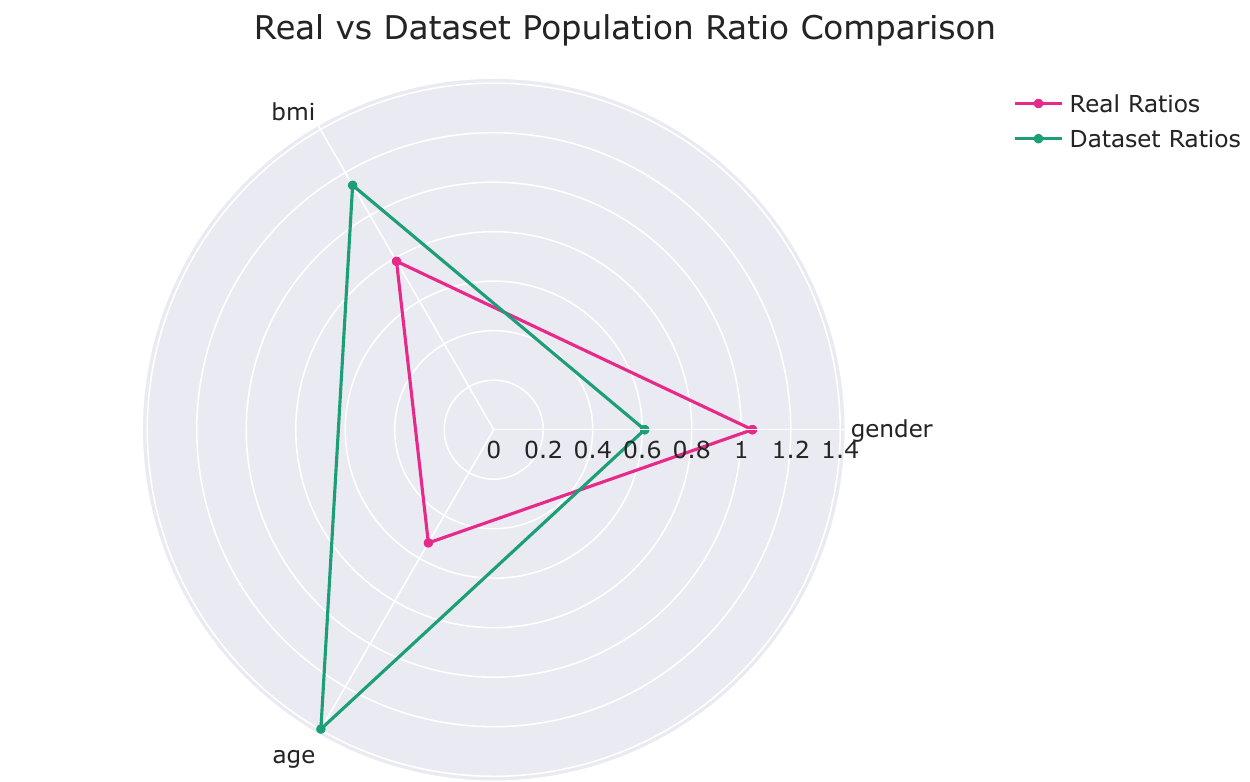}  
    \caption{Misrepresented Populations\label{l1}}
\end{subfigure}
\begin{subfigure}{.45\textwidth}
    \centering
    \includegraphics[trim={0.3cm 0 0 0},clip, width=1\linewidth]{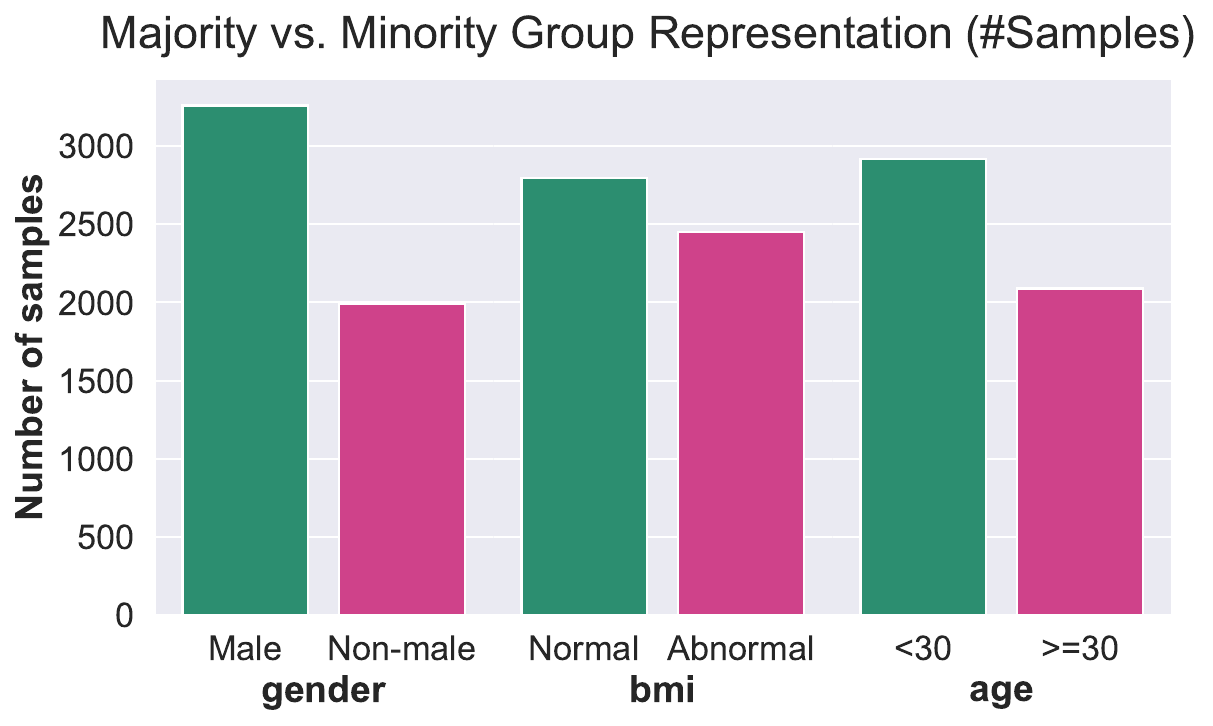}  
    \caption{Underrepresented Populations\label{l2}}
\end{subfigure}
\hfill
\begin{subfigure}{.4\textwidth}
    \centering
    \includegraphics[width=\linewidth]{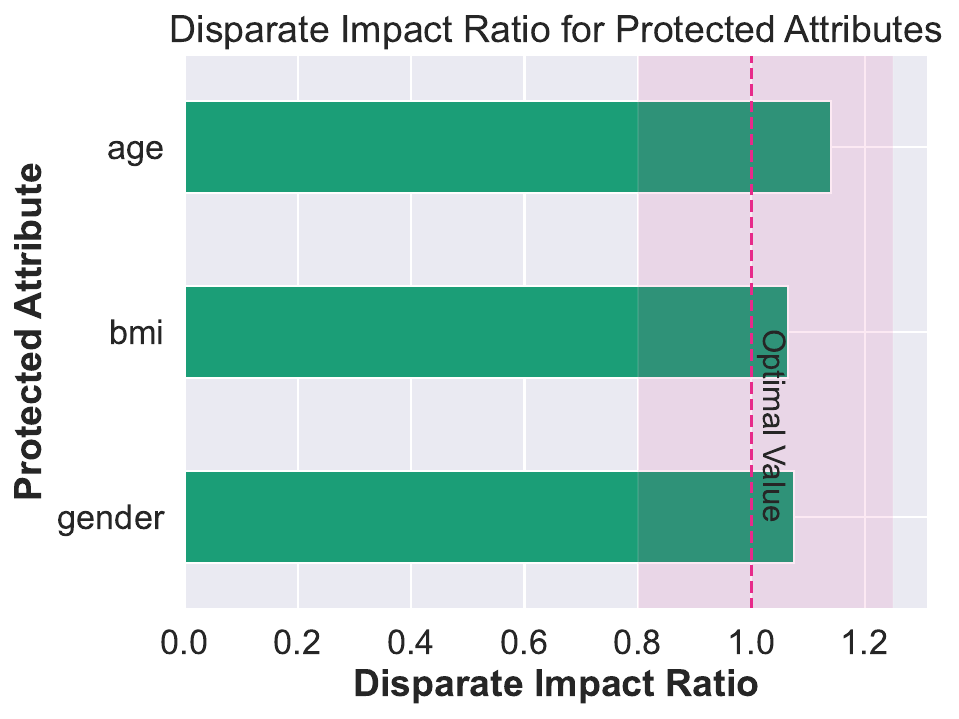} 
    \caption{Unevenly Sampled Populations\label{l3}}
\end{subfigure}
\begin{subfigure}{.45\textwidth}
    \centering
    \includegraphics[width=\linewidth]{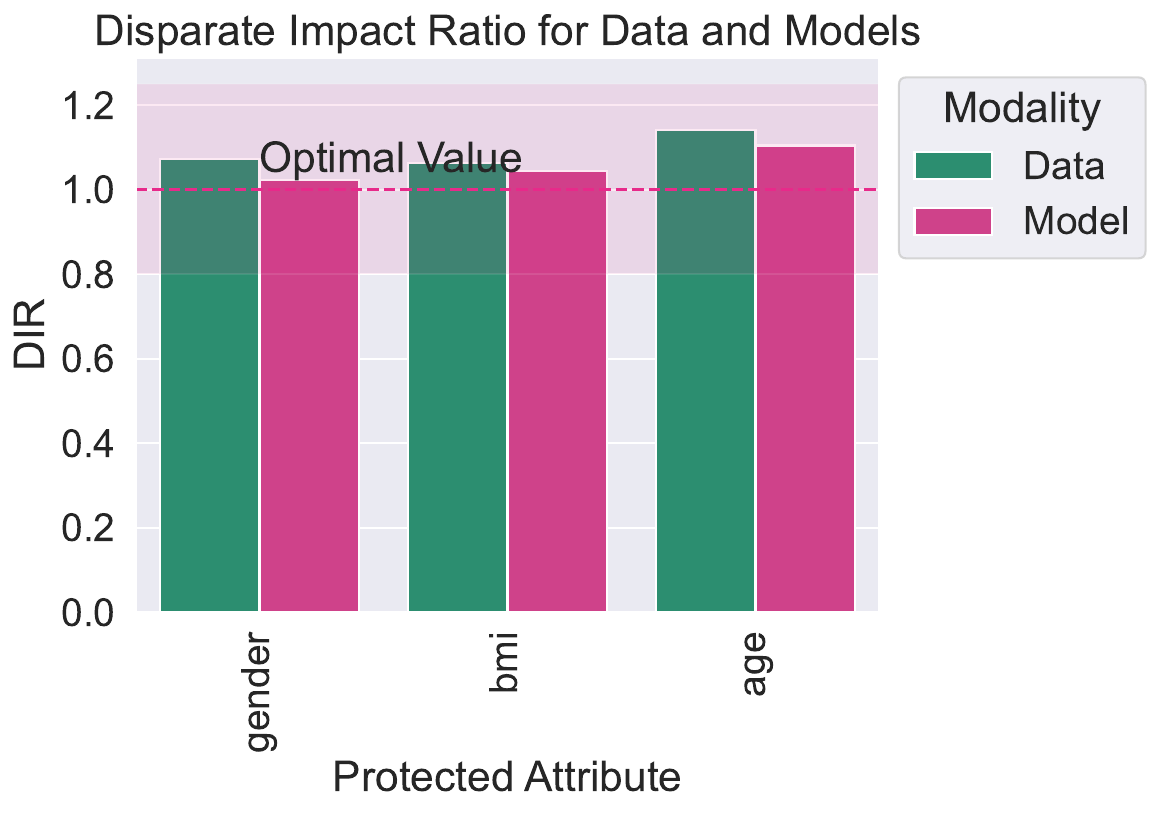} 
    \caption{\rev{Data vs. Model Biases}\label{l4}}
\end{subfigure}
\caption{\rev{LifeSnaps biases in the data generation and model building and implementation streams.}}
\label{fig:lifesnaps}
\end{figure}
\begin{figure}[t]
\centering
\begin{subfigure}{.43\textwidth}
    \centering
    \includegraphics[width=\linewidth]{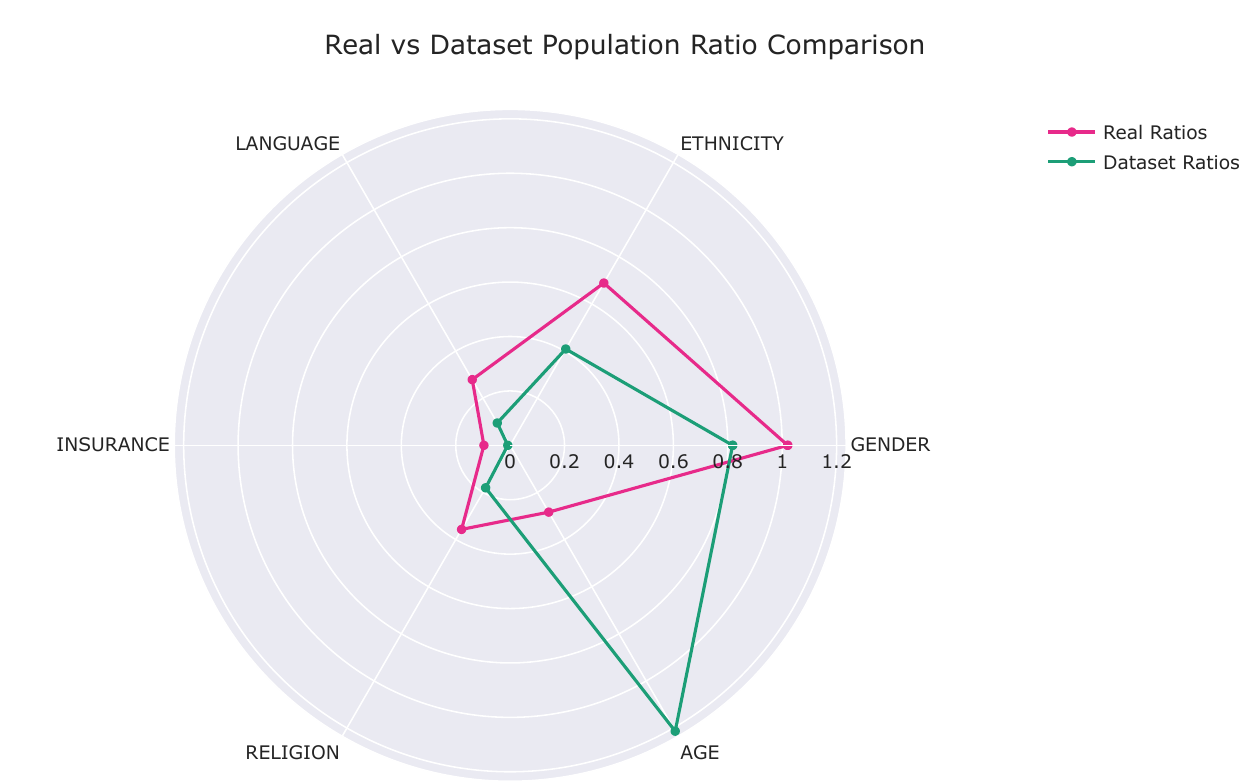}  
    \caption{Misrepresented Populations\label{m1}}
\end{subfigure}
\hspace{-11mm}
\begin{subfigure}{.56\textwidth}
    \centering
    \includegraphics[trim={0.3cm 0 0 0},clip, width=1\linewidth]{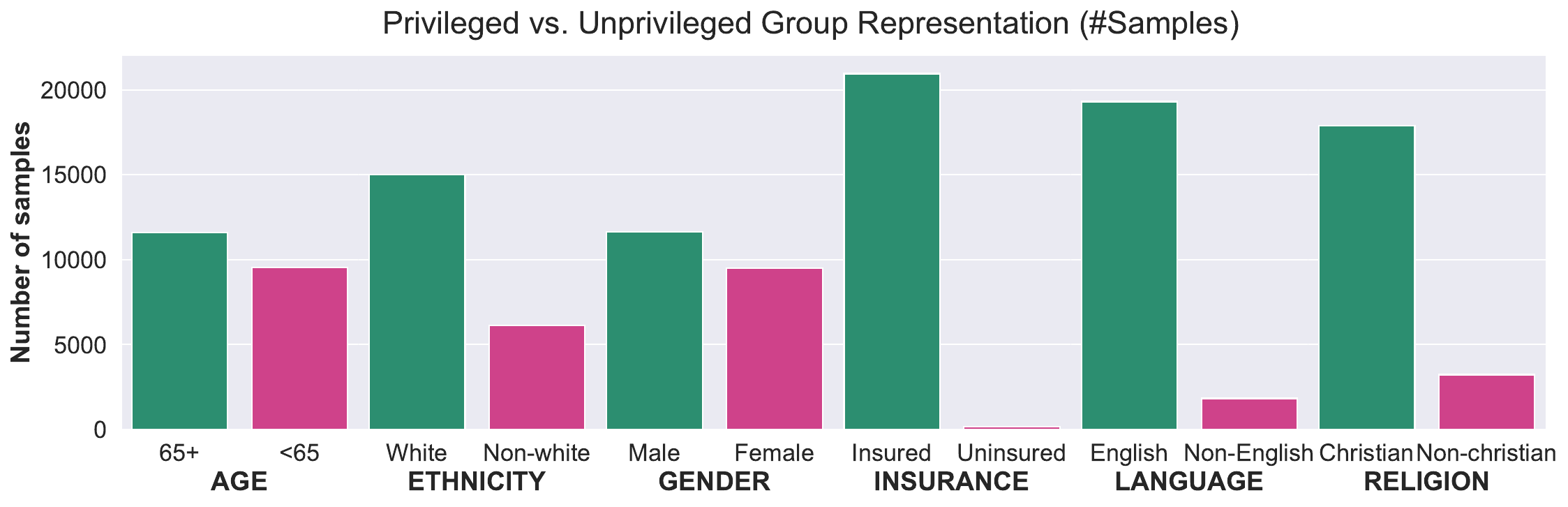}
    \caption{Underrepresented Populations\label{m2}}
\end{subfigure}
\hfill
\begin{subfigure}{.45\textwidth}
    \centering
    \includegraphics[width=\linewidth]{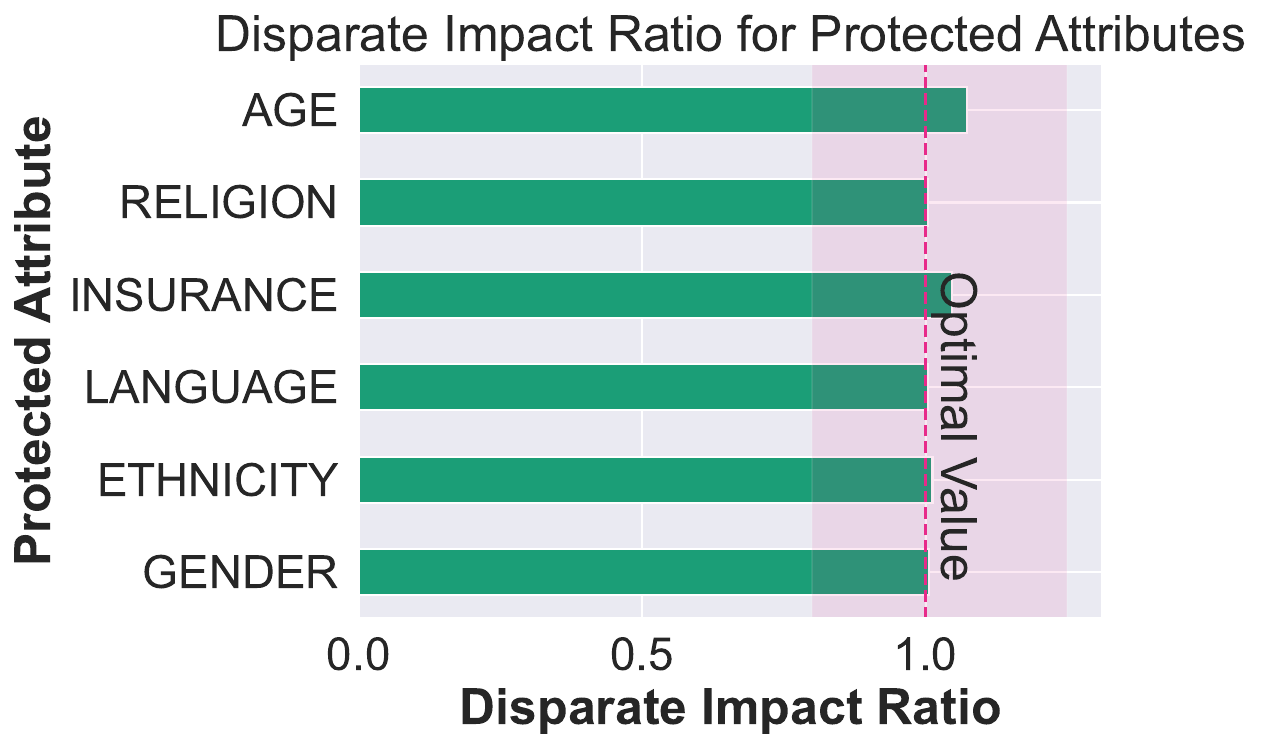} 
    \caption{Unevenly Sampled Populations\label{m3}}
\end{subfigure}
\begin{subfigure}{.4\textwidth}
    \centering
    \includegraphics[width=\linewidth]{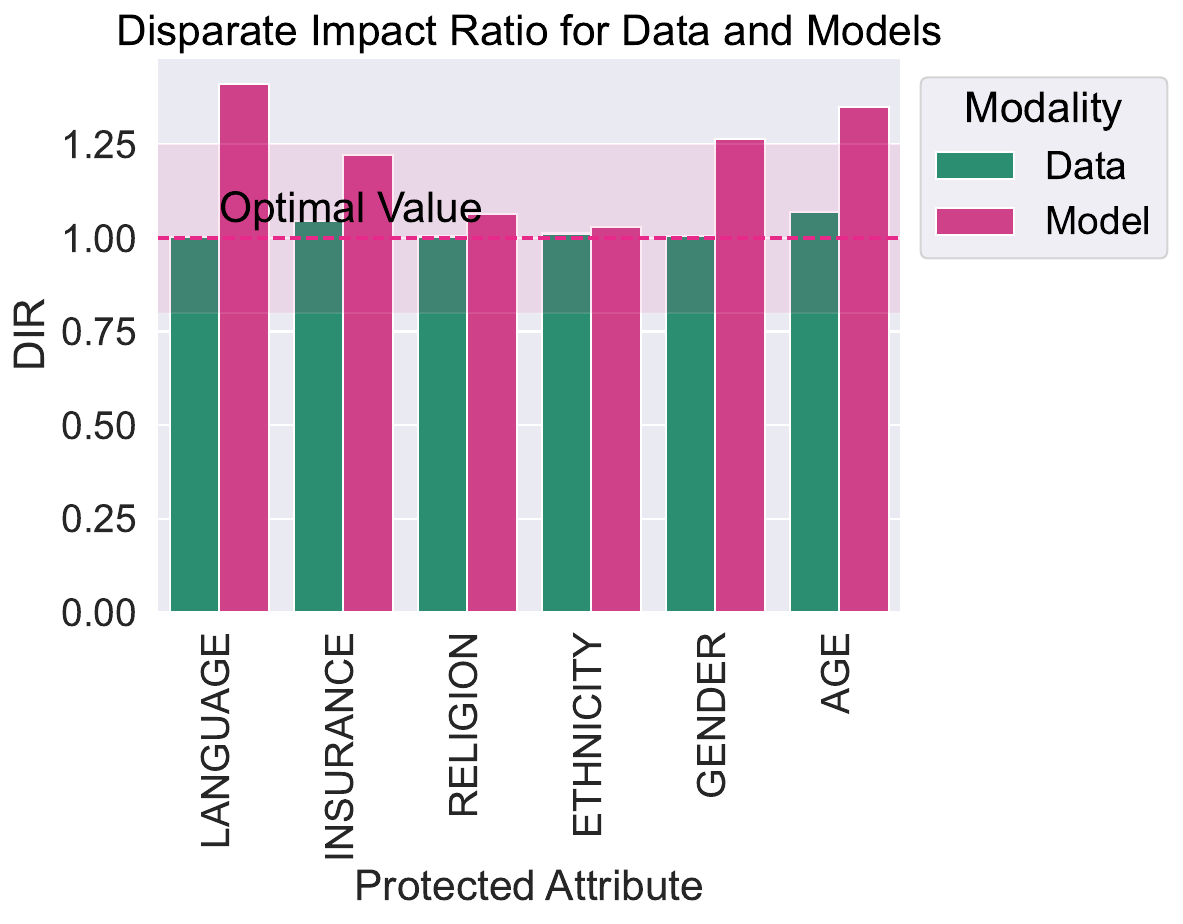} 
    \caption{\rev{Data vs. Model Biases}\label{m4}}
\end{subfigure}
\caption{\rev{MIMIC-III biases in data generation and model building and implementation streams. Data biases are amplified by deep learning models.}}
\label{fig:mimic}
\end{figure}
In exploring biases, we identified both commonalities and differences across PI datasets. Regarding the data generation stream, \textit{representation biases seem to be the norm in PI datasets}, naturally leading to \textit{learning and aggregation biases} in the model building and implementation stream and highlighting the need for increased awareness among researchers and practitioners in the field. Having said that, the identified biases are distinct in each dataset, emerging mostly from their recruitment methodology and the availability of protected attributes. 

\subsubsection*{\textbf{Bias in Rows Commonalities}} \rev{Both} datasets\rev{, similarly to MyHeart Counts,} suffer from some type of ``bias in rows'' as seen in Figures \ref{l1} and \ref{m1}. Specifically, LifeSnaps and MIMIC-III suffer from \textit{misrepresented populations}. In LifeSnaps (Figures \rev{\ref{l1} and \ref{l2}}), younger people are overrepresented due to university-based recruitment, while in MIMIC-III (Figures \rev{\ref{m1} and \ref{m2}}) older people are overrepresented due to ICU-based recruitment. Additionally, while gender and ethnicity representation is improved compared to MyHeart Counts, still white males are overrepresented in all three datasets. 
MIMIC-III, similarly to MyHeartCounts, suffers from \textit{underrepresented populations}, such as uninsured, non-white, non-English-speaking, or non-christian users (Figure~\ref{m2}). These biases are, in turn, propagated to the baseline learning models (Figure~\ref{m4}), in line with prior work \cite{roosli2022peeking}.
\subsubsection*{\textbf{Bias in Columns Differences}} When ``bias in columns'' is explored, contrary to the MyHeart Counts data, both datasets are evenly sampled in terms of outcome labels, namely physical activity in LifeSnaps and in-hospital mortality for MIMIC-III (Figures \ref{l3} and \ref{m3}, respectively). Hence, these findings may not directly generalize to other PI datasets but are still included in our methodology for completeness and visibility. Specifically, contrary to population demographics which can capture misrepresented and underrepresented groups, an analysis for unevenly sampled populations is not commonly performed during data exploration, whereas in certain cases, such as MyHeart Counts, it could reveal behavioral discrepancies across populations.
\subsubsection*{\textbf{\rev{Model Bias}}} \rev{In line with prior research \cite{roosli2022peeking}, we encounter biases in MIMIC's deep learning benchmark models. Specifically, as seen in Figure~\ref{m4}, despite even data sampling, models amplify biases for non-English-speaking, uninsured, female, and elderly users. Similar trends, independent of demographics, were noticed in the MyHeart Counts dataset. On the contrary, we do not notice any statistically significant differences in the LifeSnaps dataset (Figure~\ref{l4}), possibly due to its limited sample size compared to the other two datasets.}

Overall, these findings concerning generalizability highlight the need for comprehensive data and model evaluation in PI and, by extension, mHealth. It is high time PI researchers and practitioners looked beyond performance-only metrics to human-centric metrics capturing biases and demographic parity.   

\section{\rev{Discussion}}\label{discussion}
\rev{Unacknowledged data and model biases in PI applications for health and well-being can have critical personal and societal consequences; they can perpetuate health disparities~\cite{mamykina2015adopting}, result in misdiagnoses or delayed diagnoses for certain population segments \cite{obermeyer2019dissecting}, affect resource allocation and access to healthcare services \cite{sjoding2020racial}, or reinforce stereotypes and stigma \cite{lupton2014self}. This section discusses our findings concerning biases in PI systems and provides guidelines on mitigating identified biases in their ML life cycle (\S\ref{implications}). It also delineates the limitations of our work and areas for future research (\S\ref{limitations}).}
\subsection{\rev{Findings \& Implications}}\label{implications}

\subsubsection*{\rev{Data Generation Stream}}
As illustrated by our findings, pre-existing \textit{historical biases} are present in digital biomarkers, due to well-documented phenomena, such as the global inequality in physical activity and the digital divide, leading to data generation that is not representative of the general population. \rev{This is indeed} the case in the MyHeart Counts dataset, where female, non-white, underweight, overweight or obese, young, and hypertensive users, are undersampled. \rev{Unacknowledged historical biases, though, can creep into the ML pipeline perpetuating social injustices.} Yet, even within well-sampled user groups, data imbalances, either in terms of user attributes or measured behaviors, are still prevalent due to realistic differences across user segments. 
Specifically, in PI, we see significant underrepresentation of minority groups across all protected attributes and measured behavioral differences ---not necessarily realistic--- for users with diabetes, joint issues, unhealthy BMI, non-white users, and females. \rev{Unresolved representation biases can lead to performance discrepancies for minority groups, which in turn might lead to differences in treatment or care \cite{gottlieb2022assessment}.}
Finally, PI data is susceptible to \textit{measurement biases}, due to the heterogeneity in input modalities, performance and hardware differences across generations of devices, and usage of third-multiple party apps of unknown accuracy. Females are especially affected by such biases in our dataset, as they tend to own older devices with fewer capabilities and use more fitness-related third-party apps. \rev{Unknown measurement biases in seemingly ``objective'' sensor data can lead to errors in downstream tasks that disproportionally affect certain protected groups.} 

\rev{In an initial attempt to offer guidance to researchers in the field of ubiquitous computing, we present the following guidelines in the context of the data generation stream on how to mitigate the impact of historical, representation, and measurement biases, respectively:}

    \begin{mybox1}
        \rev{\textbf{Guideline \#1:} To identify historical biases relevant to the use case at hand, consult prior literature and domain experts (e.g., oximeters are proven susceptible to biases against darker skin tones \cite{gottlieb2022assessment}) or conduct small-scale feasibility studies with relevant and diverse demographics.}
        \smallskip
        
        \rev{\textbf{Guideline \#2:} If data are self-collected, aim for diverse user recruitment and collect and report relevant protected attributes (e.g., via datasheets for datasets and data statements). Otherwise, evaluate algorithms in generalizable cross-dataset benchmarks \cite{xu2023globem} and inclusive synthetic data \cite{van2021decaf}, whenever possible. In either case, consider appropriate data manipulation actions to alleviate biases, e.g., re-sampling/re-balancing populations conditioned on demographic attributes.}
        \smallskip
        
        \rev{\textbf{Guideline \#3:} When working with data originating from diverse devices, investigate device ownership differences conditioned on demographic attributes. Also, incorporate uncertainty estimation approaches and be transparent about possible measurement error effects in downstream tasks.}
    \end{mybox1}

    \rev{Summing up, we believe that our findings shed light on the biases that can creep into the data generation and model building, and implementation streams of PI technologies. While our mitigation guidelines are by no means exhaustive, they provide a starting point for researchers and practitioners to incorporate bias assessments ``by design'' in the life cycle of their works to alleviate the potential negative effects of such biases.}

\subsubsection*{\rev{Model Building and Implementation Stream}}
Based on our results, digital biomarkers representation biases can be propagated or even amplified by learning models, regardless of the inclusion of protected attributes in the feature set\rev{. This is} due to the existence of proxy variables in PI data that can be used by models to infer hidden protected attributes. Such \textit{aggregation biases} are also prevalent in our use case for users with joint issues, diabetes, hypertension, and female users. \rev{These biases may (or may not) lead to discrimination depending on the context. Yet, mitigating them is the safest way to ensure fairness.} Additionally, common learning choices in PI, such as personalization, can introduce \textit{learning biases}, if trained on biased data. \rev{In our case, they perform worse -in terms of bias- across all attributes, while, in extreme cases (e.g., diabetic users), they can even introduce maximum bias.} \rev{Accuracy gains emerging from alternative learning choices can be tempting, but their trade-offs should be thoroughly assessed. On a different note,} our empirical results illustrate that model performance is highly susceptible to the representativeness of the PI benchmark used and highlight how \textit{evaluation biases} can affect ubiquitous models in the evaluation phase. \rev{In such cases, performance and fairness drift can emerge if the evaluation data is not representative of the target population.} 
\rev{Finally, }the application of ML in PI is not free of \textit{deployment biases}, which can emerge from outdated evaluation practices emerging from the PI systems' early applications or the false assumption of autonomous PI systems' existence. \rev{Yet, mischoosing evaluation metrics or focusing solely on aggregate metrics can hide discrepancies in performance for minority groups. Developing high-stakes systems in isolation might also lead to (unintended) system misuse.} 

\rev{To provide guidance in the context of model building and implementations, we offer guidelines to alleviate the potential negative effects of the aggregation, learning, evaluation, and deployment biases, respectively:}

    \begin{mybox1}
        \rev{\textbf{Guideline \#4:} Utilize fairness toolkits, such as FairLearn, AIF360, and Aequitas for implementations of pre-, in-, and post-processing bias mitigation algorithms and fairness metrics.}
        \smallskip
        
        \rev{\textbf{Guideline \#5:} Move beyond accuracy in evaluating learning paradigms by incorporating fairness metrics in the evaluation pipeline of ML models, conditioning performance on intersections of protected attributes.}\smallskip

        \rev{\textbf{Guideline \#6:} Aim for representative and realistic evaluation datasets ---beyond carefully-curated benchmarks---, if available, or reassess your model after deployment. Re-train with the target population's data if you encounter performance drifts conditioned on demographic attributes.}
        \smallskip
        
        \rev{\textbf{Guideline \#7:} Choose multiple, appropriate fairness metrics based on ``fairness trees'' \cite{saleiro2018aequitas} and domain expertise for the use case at hand. Consider a human-in-the-loop design approach for high-stakes applications to account for human biases that affect system design.}
    \end{mybox1}

\subsection{Limitations \& Future Work\label{limitations}}
\rev{While MyHeart Counts offers scale and access to protected attributes, as outlined in \S\ref{configuration}, and our generalizability study supports our findings as described in \S\ref{generalizability}, additional analyses may be necessary to fully comprehend bias in PI.
For instance, certain protected attributes in the dataset have incomplete categorization, e.g., gender is treated as a binary concept, while others might be fully absent regardless of relevance to the use case, e.g., physical or mental disability and pregnancy. At the same time, the selected dataset is US-based, not capturing activity patterns across the global population. Hence our findings might not be directly applicable across protected attributes and geographical contexts. Finally, while activity tracking is the most common functionality in PI systems \mrev{and prior work has highlighted worldwide physical activity inequality \cite{althoff2017large}}, our dataset does not capture more advanced health features, such as heart monitoring and fertility tracking. Still, it is important to recognize that different use cases might incorporate different biases. Hence, while our findings shed light on the previously unexplored field of PI biases, they should be further corroborated across different contexts, such as demographics, geographical regions, and use cases.}

Appropriate PI datasets for fueling future fairness research in the domain are still lacking. Due to the sensitivity of the data at hand, many datasets are proprietary with restrictive Institutional Review Board (IRB) agreements, \rev{but inclusive, open datasets could significantly advance the domain}. Also, given the prevalence of small-scale datasets, future work should focus on quantifying biases in small digital biomarkers data, as realistically, most institutions will never acquire big data \cite{baeza2018big}. Additionally, due to closed-sourced data and algorithms, there is a lack of established benchmarks, especially regarding emerging PI tasks, such as fertility prediction, or AFib detection. To this end, similarly to the work of \citet{harutyunyan2019multitask}, future work should create inclusive and representative benchmarks for tasks within the PI domain. Beyond that, there is work to be done in quantifying \rev{and mitigating} bias in sequential physiological and behavioral data. For instance, many PI tasks are formulated as regression problems, but regression-specific fairness metrics \rev{and mitigation approaches} are limited in the literature \cite{gursoy2022error,agarwal2019fair}. 
Finally, due to privacy considerations for sensitive digital biomarkers, many times PI data are not accompanied by protected attributes for the population they describe, making it cumbersome to perform a \rev{bias and fairness} evaluation. To this end, future work should investigate the space of ``fairness in unawareness'', or, in other words, how you can quantify and mitigate biases in the absence of protected attributes.

\section{Conclusions}\label{conclusions}
This paper presents the first-of-its-kind, \rev{comprehensive} study of bias in PI by analyzing the most extensive digital biomarkers data to date. In response to our \rev{RQs}, we show that bias exists across all stages of the life cycle, both in the data generation and model building and implementation streams.  Different user minorities are affected by diverse types of bias, but users with diabetes, joint issues, or hypertension and female users show higher degrees of impact adversity in our MyHeart Counts use case due to representation, aggregation, and learning biases.
Our findings echo concerns similar to those raised in the evaluation for healthcare technologies \cite{ahmad2020fairness}. While some of our findings are specific to the investigated use case, they can \rev{mostly} be extended to \rev{other} PI tasks.

\begin{acks}
This project has received funding from the European Union’s Horizon 2020 research and innovation programme under the Marie Skłodowska-Curie grant agreement No 813162. The content of this paper reflects only the authors' view and the Agency and the Commission are not responsible for any use that may be made of the information it contains. Results presented in this work have been produced using the Aristotle University of Thessaloniki Compute Infrastructure and Resources. The authors would like to acknowledge the support provided by the Scientific Computing Office throughout the progress of this research work.
\end{acks}

\bibliographystyle{ACM-Reference-Format}
\bibliography{tidy-base}

\pagebreak
\appendix
\rev{\section{Model Architectures\label{ap:modelArchitecture}}}
\noindent\textbf{\rev{Baseline Model.}}
\rev{For our baseline model,} we consider the following setting: we are given a time-series dataset $S=\{S^{G,0}, S^{G,1}\}$ of users segmented into two groups, $G0$ and $G1$, based on protected attribute $G$ (e.g., gender, age, etc.). The user data within each group are denoted as $S^{G,g}=\{s_1^{G,g},\ldots,s_K^{G,g}\}$, where $g=\{0,1\}$ and K is the number of users per group, conditioned on protected attribute $G$. Furthermore, the data of each user are stored as $s_i=\{X_i,y_i\}$, where input time series (step count values) of users $i=1,\ldots,K$, are stored in $X_i\in \mathbb{R}^{D_x x 1}$, where $D_x=48$ (unaware model) is the length (in time steps) of a sample daily activity in the data, or $D_x=56$ (aware model) is the length of a sample daily activity in the data plus the protected attribute features.
Formally, our deep neural network architecture receives as input the users' daily activity samples ($X$) and passes them through LSTM layers with parameters $\theta_l=\{W_l,b_l\}$, weight matrix, and bias, respectively, for each layer $l$, to produce the output $\hat{y}$.
The optimization of the network parameters for LSTM layers is obtained by minimizing the binary cross entropy loss $\alpha_{c}$ defined as:
\begin{gather*} 
\Omega^{*}=\underset{\Omega=\left\{\theta_{1}, \ldots, \theta_{3}\right\}}{\arg \min } \alpha_{c}(\hat{y}, y)=\underset{\Omega=\left\{\theta_{1}, \ldots, \theta_{3}\right\}}{\arg \min } -\frac{1}{N} \sum_{i=1}^{N}\Bigl( y_i\log(\hat{y}_i)+(1-y_i)\log(1-\hat{y}_i) \Bigl)
\end{gather*} 
where $N$ represents the number of training samples \textit{from both datasets $\{S^{G,0}, S^{G,1}\}$.}

\smallskip
\noindent\textbf{\rev{Personalized Model.}}
\rev{For our personalized model,} formally, the learning during the fine-tuning process is attained through the last layer in the network, one for the minority and one for the majority user group. Before further optimization, the group-specific layers are initialized as $\theta_4^{G,0}\leftarrow\theta_4$ and $\theta_4^{G,1}\leftarrow\theta_4$, and then fine-tuned using the data from $G0$ ($S^{G,0}$) and $G1$ ($S^{G,1}$), respectively, \textit{for each protected attribute} $G$ as:
\begin{gather*}
   \left(\theta_{4}^{G,c}\right)^{*}=\underset{\theta_{4}}{\arg \min } -\frac{1}{N} \sum_{i=1}^{N \in S^{G,c}}\Bigl( y_i\log(\hat{y}_i)+(1-y_i)\log(1-\hat{y}_i) \Bigl), \quad c=\{0,1\} \text{ and}\\
   G=\{\text{gender, ethnicity, age, bmi, heart condition, hypertension, joint problem, diabetes}\} 
\end{gather*}

The final network weights, $\theta_l=\{W_l,b_l\}$, are then used to perform the group-specific inference of next-day physical activity level from past behavior per protected attribute. 
\bigskip

\section{Fairness Taxonomy in Machine Learning\label{ap:definitions}}
Viewed through the lens of quantitative science, ML research has broadly grouped fairness into two categories: \textit{individual fairness} and \textit{group fairness}. In the broad sense, group fairness partitions the general population into groups based on sensitive (a.k.a. protected) attributes and seeks statistical equality across groups. On the other hand, individual fairness seeks for similar individuals to be treated similarly \cite{dwork2012fairness,mulligan2019thing}. 

In individual fairness, determining whether individuals are similar requires first defining what features are relevant to fairness \cite{fleisher2021s}. However, in PI, it would be incomplete to define such similarity solely based on digital behavioral biomarkers, such as steps, or heart rate, as hidden contextual information might be significantly more relevant. \rev{To this end, we proceed with group fairness from now on.} This decision translates to our use case as exploring significant differences in allocation, representation, or error rates regarding future step goals across different population segments. For example, do females get systematically lower step goals than \rev{males}?

\begin{figure}[t]
  \centering
  \includegraphics[width=.9\linewidth]{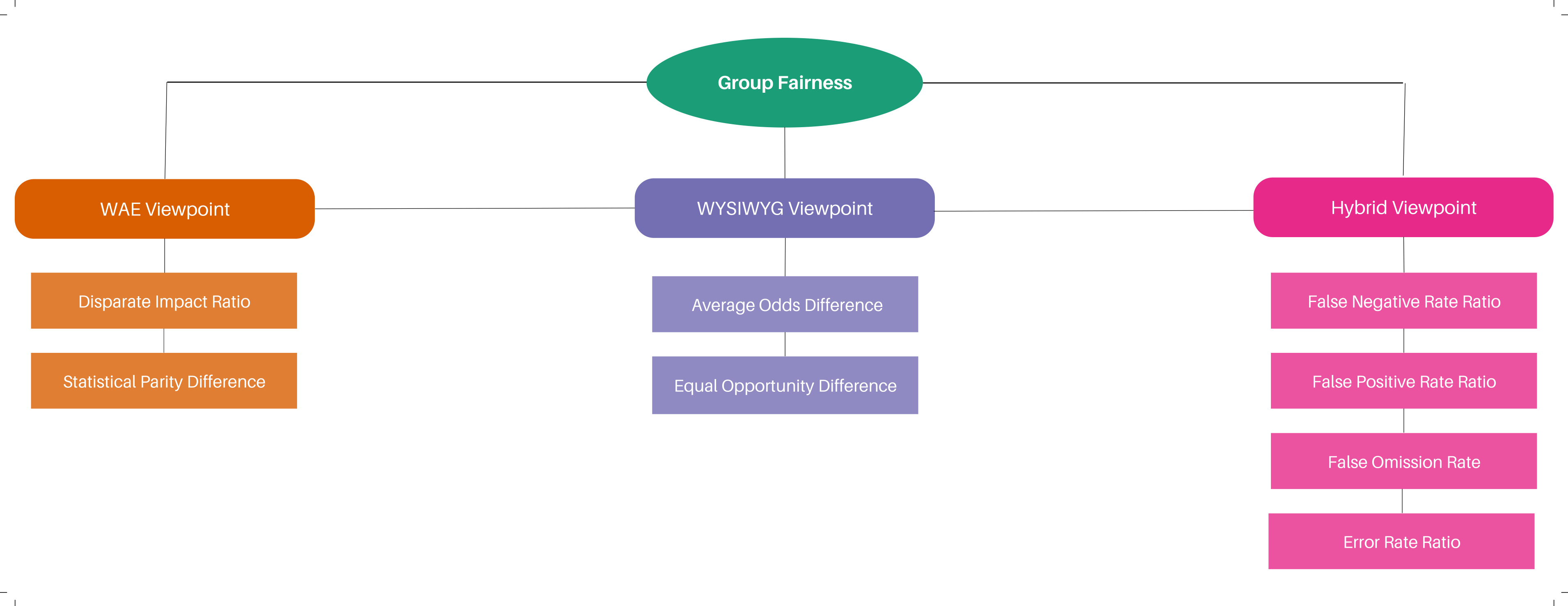}
  \caption{An organization of fairness metrics in ML according to different worldviews: WAE, WYSIWYG, and hybrid.\label{fig:biasMeasures}}
\end{figure}

Within group fairness, there are still two opposing viewpoints: we’re all equal (WAE) and what you see is what you get (WYSIWYG) \cite{friedler2021possibility,yeom2021avoiding}. The WAE viewpoint considers that all groups have similar abilities to perform the task, e.g., all groups of people are equally capable of walking more, while the WYSIWYG viewpoint holds that the data reflect each group's ability to perform the task, e.g., some groups of people might be less capable. 

\begin{table}[t]
\caption{The confusion matrix for performance measurement of models, including standard metrics.}
\label{tab:cm}
\renewcommand{\arraystretch}{2}
\begin{tabular}{lllll}
\cline{3-4}
 &  & \multicolumn{2}{c}{\textbf{True}} &  \\ \cline{3-4}
 &  & \multicolumn{1}{l}{Positive Label} & Negative Label &  \\ \hline
\multicolumn{1}{c}{\multirow{2}{*}{\textbf{Predicted}}} & Positive Label & \multicolumn{1}{l}{TP} & FP & \multicolumn{1}{l}{\begin{tabular}[c]{@{}l@{}}False Discovery Rate\\ FDR = $\frac{FP}{FP+TP}$\end{tabular}} \\ \cline{2-5} 
\multicolumn{1}{c}{} & Negative Label & \multicolumn{1}{l}{FN} & TN & \multicolumn{1}{l}{\begin{tabular}[c]{@{}l@{}}False Omission Rate\\ FOR = $\frac{FN}{FN+TN}$\end{tabular}} \\ \hline
 &  & \multicolumn{1}{l}{\begin{tabular}[c]{@{}l@{}}False Negative Rate\\(a.k.a Miss-rate)\\FNR = $\frac{FN}{TP+FN}$\end{tabular}} & \begin{tabular}[c]{@{}l@{}}True Negative Rate\\(a.k.a. Specificity)\\TNR=$\frac{TN}{FP+TN}$\end{tabular} &  \\ \cline{3-4}
 &  & \multicolumn{1}{l}{\begin{tabular}[c]{@{}l@{}}True Positive Rate\\(a.k.a. Sensitivity)\\TPR = $\frac{TP}{TP+FN}$\end{tabular}} & \begin{tabular}[c]{@{}l@{}}False Positive Rate\\(a.k.a. Fall-out)\\FPR = $\frac{FP}{FP+TN}$\end{tabular} &  \\ \cline{3-4}
\end{tabular}
 \renewcommand{\arraystretch}{1}
\end{table}
Overall, every metric tries to quantify the difference in performance between privileged and unprivileged users. 
\rev{Here, we} present the most common fairness metrics across all viewpoints, namely WAE, WYSIWYG, and hybrid \rev{(Figure~\ref{fig:biasMeasures})}. The confusion matrix (see Table~\ref{tab:cm}) is the heart of \rev{ML} performance measurement and is also used in the fairness metrics definitions below.  All metrics are expressed as \textit{ratios} or \textit{differences} between unprivileged (u) and privileged (p) groups. Note that $D$ is the user sample, $\hat{Y}$ is the predicted label, and ``$\text{pos\_label}$'' is \rev{the} favorable outcome scenario (e.g., high physical activity). 

\smallskip
\noindent\textbf{\rev{WAE Metrics.}} The WAE viewpoint supports that data, e.g., step counts, may contain biases, so \rev{the distribution differences} across groups should not be mistaken for a difference in ability. The two most commonly used WAE metrics are the \textit{Disparate Impact Ratio (DIR)} and the \textit{Statistical Parity Difference (SPD)} (Table~\ref{tab:WAEMetrics}).

\begin{table}[t]
\caption{WAE Metrics' definitions, formulas, and task and bias interpretations specific to our use case.}
\label{tab:WAEMetrics}
\resizebox{\columnwidth}{!}{%
\renewcommand{\arraystretch}{2.5}
\begin{tabular}{lp{0.2\textwidth}p{0.25\textwidth}p{0.20\textwidth}p{0.2\textwidth}}
\hline
\textbf{Metric} & \textbf{Definition} & \textbf{Formula} & \textbf{Task Interpretation} & \textbf{Bias Interpretation} \\ \hline
DIR & The ratio of base or selection rates between unprivileged and privileged groups. & $\displaystyle\frac{\Pr(\hat{Y}=\text{pos\_label}\mid\text{D}=\text{u})}{\Pr(\hat{Y} = \text{pos\_label}\mid\text{D}=\text{p})}$ & How many users receive high activity goals in the unprivileged group compared to the privileged group? & A low $DIR$ ($DIR<1$) indicates that the unprivileged user group systematically receives fewer high activity goals. \\ \hline
SPD & The difference in selection rates between unprivileged and privileged groups. & $\displaystyle\Pr(\hat{Y}=\text{pos\_label}\mid\text{D}=\text{u})-\Pr(\hat{Y}=\text{pos\_label}\mid\text{D}=\text{p})$ & Same as above & A low $SPD$ ($DIR<0$) indicates that the unprivileged user group systematically receives fewer high activity goals. \\ \hline
\end{tabular}%
}
\end{table}

\smallskip
\noindent\textbf{\rev{Hybrid Metrics.}}
Hybrid metrics lie in-between the two viewpoints. The most commonly used hybrid metrics, depending on the problem's context, are the \textit{False Positive Rate (FPR) Ratio}, the \textit{False Negative Rate (FNR) Ratio}, the \textit{False Omission Rate (FOR) Ratio}, and the \textit{Error Rate Ratio (ERR)} (Table~\ref{tab:HybridMetrics}).

\begin{table}[t]
\caption{Hybrid Metrics' definitions, formulas, and task and bias interpretations specific to our use case.}
\label{tab:HybridMetrics}
\resizebox{\columnwidth}{!}{%
\renewcommand{\arraystretch}{2.5}
\begin{tabular}{lp{0.25\textwidth}p{0.15\textwidth}p{0.20\textwidth}p{0.25\textwidth}}
\hline
\textbf{Metric} & \textbf{Definition} & \textbf{Formula} & \textbf{Task Interpretation} & \textbf{Bias Interpretation} \\ \hline
FPR Ratio & The ratio between the number of negative outcomes wrongly categorized as positive, i.e., false positives (FP), and the total number of actual negative outcomes regardless of classification. & $\displaystyle\frac{FPR_{D = \text{u}}}{FPR_{D = \text{p}}}$ & From all the low active users, how many wrongfully received high activity goals? & A low $FPR$ Ratio ($\text{FPR Ratio}<1$) indicates that the privileged low active user group systematically receives more high activity goals compared to the unprivileged low active user group. \\ \hline
FNR Ratio & The ratio between the number of positive outcomes wrongly categorized as negative, i.e., false negatives (FN), and the total number of actual positive outcomes regardless of classification. & $\displaystyle\frac{FNR_{D = \text{u}}}{FNR_{D = \text{p}}}$ & From all the highly active users, how many wrongfully received low activity goals? & A high $FNR$ Ratio ($\text{FNR Ratio}>1$) indicates that the unprivileged highly active user group systematically receives more low activity goals compared to the privileged highly active user group. \\ \hline
FOR Ratio & The ratio between the outcomes wrongly categorized as negative, i.e., FN, and the total number of classified negative outcomes. & $\displaystyle\frac{FOR_{D = \text{u}}}{FOR_{D = \text{p}}}$ & From all the users that were given low activity goals -rightfully so or not-, how many were actually highly active? & A high $FOR$ Ratio ($\text{FOR Ratio}>1$) indicates that the unprivileged user group systematically receives more wrong low activity goals compared to the privileged user group. \\ \hline
ERR & The ratio between the erroneous outcomes, i.e., FN or FP, and the total number of outcomes. & \begin{tabular}[c]{@{}l@{}}$\displaystyle\frac{ER_{D = \text{u}}}{ER_{D = \text{p}}} \text{, where}$\\ $\displaystyle\text{ER} = \frac{FP+FN}{P+N}$\end{tabular} & How many times was the activity level prediction wrong? & A high $ERR$ ($\text{ERR}>1$) indicates that the unprivileged user group systematically receives more wrong goals -low or high- compared to the privileged user group. \\ \hline
\end{tabular}%
}
\end{table}

\smallskip
\noindent\textbf{\rev{WYSIWYG Metrics.}}
The WYSIWYG viewpoint supports that data correlates well with future \rev{behavior} and that \rev{they can be used} to compare the \rev{users' abilities}. The two most commonly used WAE metrics are the \textit{Average Odds Difference (AOD)} and the \textit{Equal Opportunity Difference (EOD)}.

\begin{table}[t]
\caption{WYSIWYG Metrics' definitions, formulas, and task and bias interpretations specific to our use case.}
\label{tab:WYSIWYGMetrics}
\resizebox{\columnwidth}{!}{%
\renewcommand{\arraystretch}{2.5}
\begin{tabular}{lp{0.20\textwidth}p{0.30\textwidth}p{0.20\textwidth}p{0.25\textwidth}}
\hline
\textbf{Metric} & \textbf{Definition} & \textbf{Formula} & \textbf{Task Interpretation} & \textbf{Bias Interpretation} \\ \hline
EOD & The difference of true positive rates between the unprivileged and the privileged groups. & $\displaystyle\text{TPR}_{D = \text{u}} - \text{TPR}_{D = \text{p}}$ & From all the highly active users, how many were actually given high activity goals? & A low $EOD$ ($EOD<-0.1$) indicates that the unprivileged highly active user group systematically receives fewer high activity goals compared to the privileged highly active user group. \\ \hline
AOD & The average difference between the FPR and the TPR between unprivileged and privileged groups. & $\frac{(FPR_{D = \text{u}} - FPR_{D = \text{p}})+ (TPR_{D = \text{u}} - TPR_{D = \text{p}})}{2}$ & TBD & TBD \\ \hline
\end{tabular}%
}
\end{table}

\begin{figure}[t]
  \centering
  \includegraphics[width=.9\textwidth,height=\textheight,keepaspectratio]{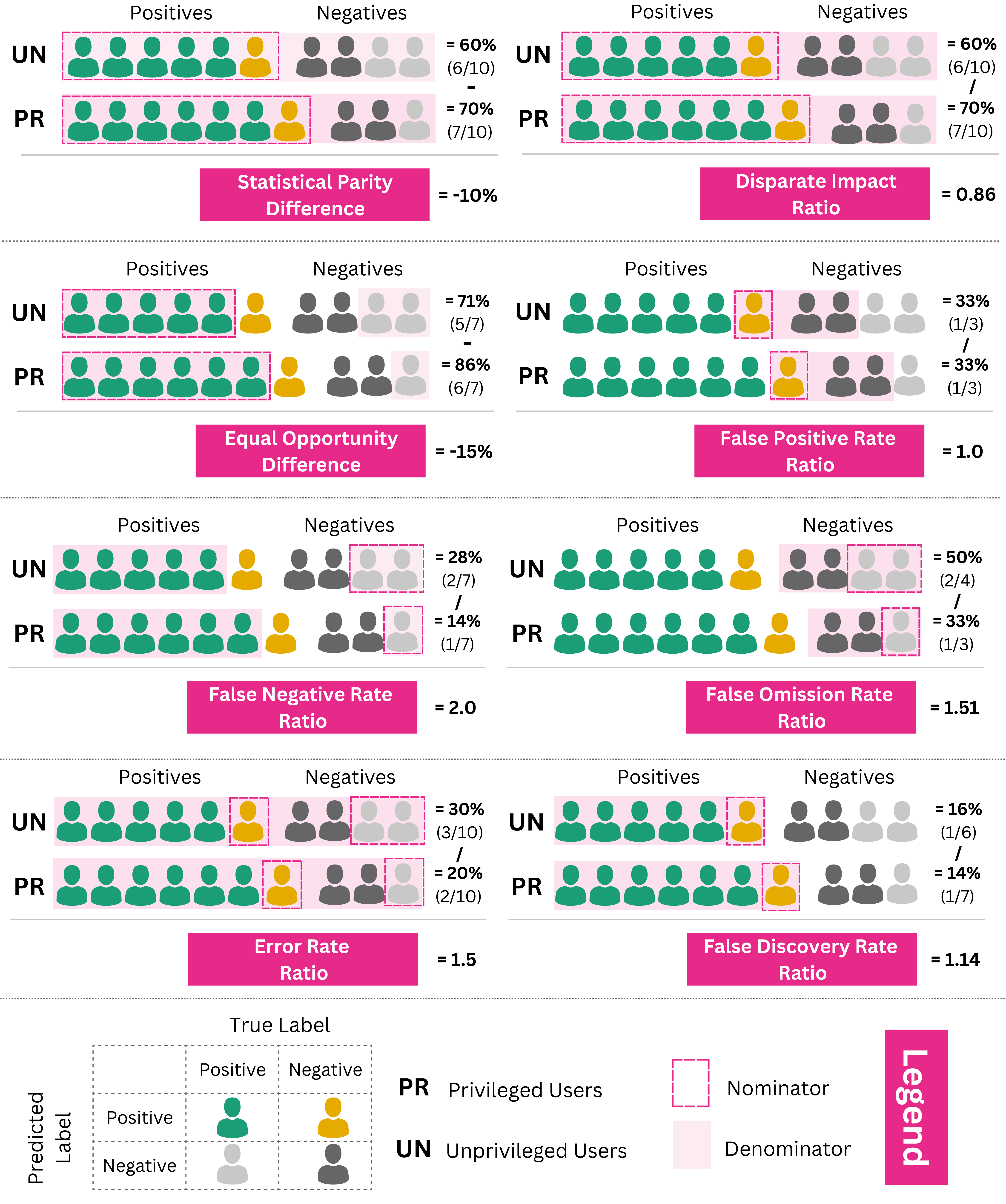}
  \caption{A graphical overview of the fairness metrics discussed.\label{fig:overviewMetrics}}
\end{figure}

\smallskip
\noindent\textbf{\rev{Fairness Metrics' Value Ranges.}}
Figure~\ref{fig:overviewMetrics} gives a graphical overview of all fairness metrics discussed. We notice that difference-based metrics have a different range of values compared to ratio-based metrics. Specifically, difference-based metrics are within the $[-100\%,+100\%]$ range, while ratio-based metrics are within the $[0,\infty]$ range. For both categories, a value of 1.0 is optimal, indicating demographic parity. Anything greater or less than the optimal value indicates some level of bias. According to AIF360, accepted difference-based metrics' values are within the range $[-0.1,+0,1]$, and accepted ratio-based metrics' values are within [0.8,1.25], but such ranges are not universally accepted and might be adjusted on a task-by-task basis. Table~\ref{tab:metricsValues} indicates which user group, namely privileged or unprivileged, benefits based on a group fairness metric's value.
\begin{table}[t]
    \caption{Value interpretation for group fairness metrics. The table shows which user group is treated unfairly -in a negative manner- in each case. UN indicates the unprivileged user group, and PR indicates the privileged user group. UN $\sim$ PR indicates a fair outcome. Notice that the same values may mean different things in the case of ratio-based metrics.}
    \begin{subtable}[t]{.45\textwidth}
        \caption{Ratio-based metrics.\label{tab:metricsValues}}
        \raggedright
         \begin{tabular}{clll}
        \hline
        \multicolumn{1}{l}{} & \multicolumn{3}{c}{\textbf{Value ($v$)}} \\ \hline
        \multicolumn{1}{l|}{\textbf{Metric}} & $v<0.8$ & $0.8<v<1.25$ & $v>1.25$ \\ \hline
        \multicolumn{1}{c|}{\textit{DIR}} & UN & UN $\sim$ PR & PR \\ \hline
        \multicolumn{1}{c|}{\textit{FOR}} & PR & UN $\sim$ PR & UN \\ \hline
        \multicolumn{1}{c|}{\textit{FNR}} & PR & UN $\sim$ PR & UN \\ \hline
        \multicolumn{1}{c|}{\textit{FPR}} & UN & UN $\sim$ PR & PR \\ \hline
        \multicolumn{1}{c|}{\textit{ERR}} & PR & UN $\sim$ PR & UN \\ \hline
        \end{tabular}
    \end{subtable}%
   \begin{subtable}[t]{.45\textwidth}
        \raggedleft
        \caption{Difference-based metrics.}
        \begin{tabular}{clll}
        \hline
        \multicolumn{1}{l}{} & \multicolumn{3}{c}{\textbf{Value ($v$)}} \\ \hline
        \multicolumn{1}{l|}{\textbf{Metric}} & $v<-0.1$ & $-0.1<v<0.1$ & $v>0.1$ \\ \hline
        \multicolumn{1}{c|}{\textit{SPD}} & UN & UN $\sim$ PR & PR \\ \hline
        \multicolumn{1}{c|}{\textit{EOD}} & UN & UN $\sim$ PR & PR \\ \hline
        \multicolumn{1}{c|}{\textit{AOD}} & UN & UN $\sim$ PR & PR \\ \hline
        \end{tabular}
    \end{subtable}
\end{table}
\end{document}